\documentclass[12pt]{article}
\usepackage{epsfig,amssymb,amsmath,psfrag}
\usepackage{caption}
\usepackage{subcaption}
\usepackage{tikz}
\usepackage{graphicx}
\usepackage{hyperref}

\def \bl  {\begin{align*}}
\def \el  {\end{align*}}

\def \be  {\begin{equation}}
\def \ee  {\end{equation}}
\def \ba  {\begin{eqnarray}}
\def \ea  {\end{eqnarray}}
\def \baa {\begin{eqnarray*}}
\def \eaa {\end{eqnarray*}}
\def \bb  {\begin {thebibliography} }
\def \eb  {\end{thebibliography}}

\def \lab #1 {\label{#1}}

\newcommand{\beq}{\begin{equation}}
\newcommand{\eeq}{\end{equation}}
\newcommand{\beqa}{\begin{eqnarray}}
\newcommand{\eeqa}{\end{eqnarray}}

\def \tr {\mathop{\rm tr}\nolimits}

\def\l<{\langle}
\def\r>{\rangle}

\def\XXint#1#2#3{{\setbox0=\hbox{$#1{#2#3}{\int}$}
     \vcenter{\hbox{$#2#3$}}\kern-.5\wd0}}

\renewcommand{\title}[1]{\vbox{\center\LARGE{#1}}\vspace{5mm}}
\renewcommand{\author}[1]{\vbox{\center#1}\vspace{5mm}}

\numberwithin{equation}{section}
\begin{document}

\thispagestyle{empty}

\begin{flushright}

\end{flushright}

\hypersetup{pageanchor=false}

\vskip2.2truecm
\begin{center}
\vskip 0.2truecm {\Large\bf
{\Large BCFW recursion for light-like Wilson loop correlators}
}\\
\vskip 1truecm
{\bf J.M. Drummond\footnote{J.M.Drummond@soton.ac.uk}, M. Rochford\footnote{M.J.Rochford@soton.ac.uk}, R. Wright\footnote{Rowan.Wright@soton.ac.uk}
}
\vskip 0.4truecm
{\emph{School of Physics and Astronomy, University of Southampton, Southampton, SO17 1BJ, UK
}}

\vskip 0.4truecm

\begingroup\bf\large

\endgroup
\vspace{0mm}

\begingroup
\textit{
 }\\
\par
\endgroup

\end{center}

\vskip 0.4truecm

\vspace{0mm}

\centerline{\bf Abstract} 
We apply the idea of holomorphic linking \cite{Bullimore:2011ni} to correlators of multiple light-like loop operators in $\mathcal{N}=4$ super Yang-Mills theory. This allows us to extend the BCFW recursion relations for planar scattering amplitude integrands or single Wilson loops in the planar theory to correlators of multiple loop operators. We discover a novel term which feeds the recursion for single loops into the relation for multiple loop operators. We illustrate the recursion relations with several examples and highlight its applicability in combination with the $\bar{Q}$-equation for multiple loop operators. Finally, we again employ holomorphic linking to derive a recursion relation for colour-exact Wilson loop correlators.
\noindent

\newpage
\setcounter{page}{1}\setcounter{footnote}{0}
\tableofcontents
\newpage

\hypersetup{pageanchor=true}

\section{Introduction}

Recursion relations for scattering amplitudes have been a powerful tool in the study of perturbative quantum field theory. The well-known BCFW relations \cite{Britto:2004ap,Britto:2005fq} permit massless on-shell tree-level amplitudes to be recursively computed starting from three-point amplitudes (defined for complex momenta). The idea of \cite{Britto:2005fq} is that the tree-level on-shell amplitudes can be treated as a rational function of their momenta (or more precisely the spinor-helicity variables associated to the momenta). The amplitude can then be deformed to become a meromorphic function of an additional auxiliary complex parameter. The poles of the deformed amplitude are related to factorisation channels of the original amplitude and thus their residues are related to on-shell amplitudes with lower multiplicity. Assuming appropriate analytic behaviour of the deformed amplitude, the original undeformed amplitude can then be reconstructed as a sum over the residues.

These relations have been exploited and generalised for many applications. Of particular relevance here is the fact that they can be cast in a manifestly supersymmetric form for supersymmetric gauge theories \cite{Brandhuber:2008pf,Arkani-Hamed:2008owk}, In particular, for $\mathcal{N}=4$ super Yang-Mills theory, they involve a single superamplitude for each multiplicity \cite{Nair:1988bq}, as the on-shell multiplet is CPT self-conjugate. The supersymmetric BCFW recursion relation can then be generalised to loop integrands for $\mathcal{N}=4$ super Yang-Mills theory in the planar limit \cite{Arkani-Hamed:2010zjl}. In this sense, it provides access to all-loop information about the theory. 

Since (super) amplitudes are dual to light-like (super) Wilson loops in the planar limit of the $\mathcal{N}=4$ theory \cite{Alday:2007hr,Drummond:2007aua,Brandhuber:2007yx,Bern:2008ap,Drummond:2008aq,Mason:2010yk}, it should be possible to understand the BCFW recursion relations, both at tree-level and for the loop integrand, from the Wilson loop side of the duality. Indeed, it was shown in \cite{Bullimore:2011ni} that this is possible. At tree level, the approach relies on making use of a twistorial formulation which describes the self-dual $\mathcal{N}=4$ theory as a holomorphic Chern-Simons theory in twistor space \cite{Witten:2003nn}\footnote{The relation of Chern-Simons actions to self-dual (super) Yang-Mills theory has a long history, see e.g. \cite{Sokatchev:1995nj}.}. The Wilson loop operator is formulated in a manner directly analogous to its spacetime version as the trace of the holonomy of the gauge field around a loop consisting of a sequence of intersecting (holomorphic) lines. The supersymmetric tree-level BCFW recursion relations then beautifully arise via the Makeenko-Migdal loop equations \cite{Makeenko:1979pb} applied to these holomorphic Chern-Simons Wilson loops. The singular configurations, corresponding to factorisation channels in the amplitude formulation of the recursion relation,  correspond to self-intersecting Wilson loop configurations in this new formulation. Since these Wilson loops are formulated in holomorphic Chern-Simons theory and the variations of the loops are holomorphic, these configurations are related to `holomorphic linking'.

As also described in \cite{Bullimore:2011ni}, the Wilson loop formulation of the recursion relation can be extended to capture the all-loop integrand. The holomorphic  Chern-Simons formulation of the self-dual sector can be completed to the full $\mathcal{N}=4$ theory by adding a tower of non-local interactions of `log-det' form \cite{Boels:2006ir}. When included in the analysis of the loop equations, these interactions then generate the dependence on the loop variables in the integrand.

Here we will show that the recursion relations obtained from loop equations and holomorphic linking can be generalised to correlators of multiple super Wilson loop operators. Such correlators have recently been analysed in \cite{Drummond:2025ulh,Drummond:2026lvq,Drummond:2026gpt}, where a number of results have been obtained making use of the twistorial formulation of the $\mathcal{N}=4$ theory and the supersymmetric loop operators. Our analysis will follow closely that of \cite{Bullimore:2011ni}, although we will have to pay close attention to the large $N$ expansion as we will be interested in the connected parts of correlators of multiple loop operators, which are suppressed at large $N$ with respect to the disconnected contributions. 

The recursion relations we obtain for tree-level correlators will contain a novel type of term which feeds the results of single loop operators into those for multiple loops. In this sense, we obtain a recursion which directly extends that for a single Wilson loop in holomorphic Chern-Simons theory. We also generalise to the full $\mathcal{N}=4$ theory by including the non-local `log-det' interactions to obtain results for loop integrands. It is then possible to go beyond the planar limit and obtain a relation satisfied by colour exact multiple Wilson loop correlators. This relation expresses correlators in terms of a sum over correlators with either a lower number of cusps on the loops or correlators of a lower Grassmann degree, which thus allows for recursive computation of colour-exact Wilson loops.

The recursion relation we obtain has some direct applications beyond obtaining explicit expressions for Wilson loop correlators. In particular, it can be used to justify a general analysis of the $\bar{Q}$-equation obeyed by correlators of multiple loop operators described in \cite{Drummond:2025ulh,Drummond:2026lvq} which is a generalisation of the $\bar{Q}$-equation for Wilson loops presented in \cite{Caron-Huot:2011dec,Bullimore:2011kg}.

\section{Twistor formulation of \texorpdfstring{$\mathcal{N}=4$}{N=4} SYM and loop operators}

Recall that, perturbatively, \(\mathcal{N}=4\) Super-Yang Mills theory may be formulated in supertwistor space $\mathbb{CP}^{3|4}$ \cite{Boels:2006ir}. We use the notation $\mathcal{Z} = (Z|\chi)$ to denote a supertwistor with bosonic twistors $(Z^A)= (Z^1,Z^2,Z^3,Z^4)$ and Grassmann counterparts $(\chi^{A'}) = (\chi^1,\chi^2,\chi^3,\chi^4)$. Note that we have the projective invariance $\mathcal{Z} \cong u \mathcal{Z} = (u Z | u \chi)$ for $u \in \mathbb{C}^*$.

Firstly, we introduce a $(0,1)$-form gauge connection $\mathcal{A}(\mathcal{Z})$ whose various components are given explicitly by the expansion
\begin{align}
\label{superA}
\mathcal{A}(Z,\chi) = a(Z,\overline{Z}) &+ \chi^{A'} \tilde{\gamma}_{A'}(Z,\overline{Z}) + \frac{1}{2!}\chi^{A'}\chi^{B'} \phi_{A'B'}(Z,\overline{Z}) \\
&+ \frac{1}{3!}\epsilon_{A'B'C'D'}\chi^{A'}\chi^{B'}\chi^{C'}
\gamma^{D'}(Z,\overline{Z}) + \frac{1}{4!}\epsilon_{A'B'C'D'}\chi^{A'}\chi^{B'}\chi^{C'}\chi^{D'} g(Z,\overline{Z}). \notag 
\end{align}
Here $\{a,\tilde{\gamma},\phi,\gamma,g\}$ are all $(0,1)$-forms with degrees of homogeneity $\{0,-1,-2,-3,-4\}$ respectively under the $u$ rescaling described above. Note that $\mathcal{A}$ is holomorphic in $\chi$ both in the sense that $\mathcal{A}$ is independent of $\bar{\chi}$ and that $\partial_{\bar{\chi}} \lrcorner \mathcal{A} =0$ (i.e. there are no $d \bar\chi$ contributions).

The $\mathcal{N}=4$ theory is described using the twistor action 
\be
S[\mathcal{A}] = S_1[\mathcal{A}] + S_2[\mathcal{A}] \,,
\ee
where the self-dual theory is described by a holomorphic Chern-Simons action,
\be
S_1[\mathcal{A}] = \beta_1 \int D^{3|4} \mathcal{Z} \wedge \textrm{tr} \Bigl(\frac{1}{2}\mathcal{A} \wedge \overline{\partial}\mathcal{A} + \frac{1}{3}\mathcal{A} \wedge \mathcal{A} \wedge \mathcal{A}\Bigr)
\ee
and the measure is defined by
\be
D^{3|4}\mathcal{Z} = \frac{1}{4!}\epsilon_{ABCD}Z^A dZ^B \wedge dZ^C \wedge dZ^D \frac{1}{4!} \epsilon_{A'B'C'D'}d\chi^{A'} d\chi^{B'} d\chi^{C'} d\chi^{D'}\,.
\ee
The expansion around the self-dual theory is given by a non-local interaction term where the interactions are confined to lines $X$ in supertwistor space,
\begin{align}
S_2[\mathcal{A}] &= \beta_2 \int d^{4|8} X \log \det\bigl(\overline{\partial} + \mathcal{A} \bigr)_X \,.
\label{logdet}
\end{align}
In order to faithfully produce the spacetime \(\mathcal{N}=4\) action, \(\beta_1\) and \(\beta_2\) must obey the relation (here strictly speaking there is a sign ambiguity depending on whether we are matching to the spacetime action in Lorentzian or Euclidean signature)
\begin{equation}
  \beta_2 = -4\pi^2\beta_1^2g^2_{\rm YM} 
\end{equation}
Note that this means that there is a rescaling freedom in matching to the action, essentially equivalent to rescaling the fermions. We will later make the choice
\begin{equation}
\beta_1 = -\frac{iC_F}{4\pi^3}, \, \, \, \, \, \, \, \, \beta_2 = \frac{4g^2C_F^2}{\pi^2 N},
\end{equation}
with \(g^2 = \frac{g^2_{\rm YM}}{16\pi^2}\). This scaling is chosen to ensure a perfect match (including numerical prefactors) between planar Wilson loop expectation values computed in the twistor formalism and planar scattering amplitudes, and also to ensure the colour-exact expectation value of a Wilson loop obeys the \(\bar{Q}\)-equation as discussed in \cite{Drummond:2026lvq}. In the analysis which follows, we will leave \(\beta_1\) and \(\beta_2\) symbolic. 

We can define a line $X$ by picking two specified points $\mathcal{Z}_A$ and $\mathcal{Z}_B$ in supertwistor space and considering the set of points $s_A \mathcal{Z}_A + s_B \mathcal{Z}_B$. We can regard $X = (\mathcal{Z}_A, \mathcal{Z}_B)$ as the image of a holomorphic embedding $\mathcal{Z}_X$ of $\Sigma = \mathbb{CP}^1$ into $\mathbb{CP}^{3|4}$,
\begin{align}
    &\mathcal{Z}_{X} : \Sigma \rightarrow \mathbb{CP}^{3|4}\,, \notag \\
    &\mathcal{Z}_{X} : s \mapsto s \mathcal{Z}_A + \mathcal{Z}_B\,,
\end{align}
where we have chosen to parametrise $\mathbb{CP}^1$ with homogeneous coordinates $(s_A,s_B) = (s,1)$ and specified the map so that $s=0$ corresponds to $\mathcal{Z}_B$ and $s=\infty$ corresponds to $\mathcal{Z}_A$.

By $(\bar{\partial} + \mathcal{A})_X$ in (\ref{logdet}) we mean the pullback to $\Sigma$ of $(\bar{\partial} + \mathcal{A})$ under the holomorphic map $\mathcal{Z}_X$.
We may rewrite (\ref{logdet}) using 
\be
\log \det (\bar{\partial} + \mathcal A) = \tr \log (\bar{\partial} + \mathcal A) = \tr \log \bar{\partial} - \sum_{r=1}^\infty \frac{1}{r} \tr \bigl(-\bar{\partial}^{-1} \mathcal{A}_X\bigr)^r
\ee
and keep only terms with at least two instances of the connection $\mathcal{A}$ in order to saturate the Grassmann integration, which yields 
\be
S_2[\mathcal{A}] = -\beta_2 \int d^{4|8}X \sum_{r=2}^{\infty}\frac{1}{r} \tr \bigl(-\bar{\partial}^{-1} \mathcal{A}_X\bigr)^r\,.
\label{trlog}
\ee
Here $\bar{\partial}^{-1}$ means any representative integral operator $\bar{\partial}_{s_0}^{-1}$ acting on $(0,1)$-forms on $\Sigma$  as
\be
(\bar{\partial}_{s_0}^{-1} \omega)(s) = \int_{\Sigma} G_{s_0}(s,s') \wedge \omega(s')\,.
\label{dbarinv}
\ee
The $(1,0)$-form Green's function $G_{s_0}(s,s)$ is defined by
\be
G_{s_0}(s,s') = -\frac{1}{2\pi i} \frac{(s-s_0)ds'}{(s-s')(s'-s_0)}\,
\label{Greensfn}
\ee
and is the solution to
\be
\bar{\partial} G_{s_0}(s,s') = i \delta(s-s') d\bar{s} \wedge ds' = i\bar{\delta}(s-s') \wedge ds'\,,
\ee
obeying $G_{s_0}(s_0,s')=0$. Note that the delta function is normalised so that
\be
\int ds \wedge d\bar{s} \, \delta(s-t) = \int ds \wedge \bar{\delta}(s-t) = -i\,.
\ee

The terms in (\ref{trlog}) are then explicitly given using
\begin{align}
\tr (-\bar{\partial}^{-1} \mathcal{A}_X)^r &= \tr  \int_{\Sigma^r} G_{s_0}(s_r,s_1) \wedge \mathcal{A}_X(s_1) \wedge \ldots \wedge G_{s_0}(s_{r-1},s_r) \wedge \mathcal{A}_X(s_r) \,, \notag \\
&= \Bigl( \frac{1}{2\pi i} \Bigr)^r \tr \int_{\Sigma^r} \frac{ds_1 \wedge \mathcal{A}_X(s_1) \wedge \ldots \wedge ds_{r} \wedge \mathcal{A}_X(s_r)}{(s_r-s_1)\ldots(s_{r-1}-s_r)}
\end{align}
and we note that all dependence on $s_0$ indeed cancels.

To define a loop operator associated to a curve $C$ we first need to define a parallel transport operator corresponding to a given line.
Given a line $X$, and a $(0,1)$-form Chern-Simons connection $\mathcal{A}$, we can find a frame $H(X,s)$, smoothly varying on $\Sigma$, such that
\be
H^{-1} (\bar{\partial} + \mathcal{A})_X H = \bar{\partial}\,.
\ee
The frame obeys the differential equation
\be
(\bar{\partial} + \mathcal{A})_X H = 0\,.
\ee
The frame $H(X,s)$ is unique up to multiplication by a gauge transformation $g(X)$ (independent of $s$),
\be
H(X,s,\bar{s}) \rightarrow H(X,s) g(X)\,.
\ee
With some choice of frame $H(X,s)$, we define the parallel transport operator
\be
U_X(s,s') \equiv H(X,s) H(X,s')^{-1} 
\ee
which then obeys the differential equation
\be
(\bar{\partial} + \mathcal{A})_{X} U_X(s,s') = 0\,
\label{Ueq}
\ee
with the boundary condition that $U_X(s',s') = 1$\,. The parallel transport operator also obeys the concatenation property
\be
U_X(s,s')U_X(s',s'') = U_X(s,s'')\,.
\ee
We may formally expand $U_X(s,s')$ in terms of the integral operator $\bar{\partial}_{s'}^{-1}$ described in (\ref{dbarinv}),
\begin{align}
\label{UXss'} 
U_{X}(s,s') &= \sum_{l=0}^{\infty} \bigl(-\bar{\partial}_{s'}^{-1} \mathcal{A}_{X}\bigr)^{l}(s) \\
&= 1 - \int_{\Sigma} G_{s'}(s,s_1) \wedge \mathcal{A}_{X}(s_1) + \int_{\Sigma} G_{s'}(s,s_1) \wedge \mathcal{A}_{X}(s_1) \int_{\Sigma} G_{s'}(s_1,s_2) \wedge \mathcal{A}_{X}(s_2) + \ldots\, \notag \\
&= 1 + \frac{s-s'}{2 \pi i} \int  \frac{ds_1 \wedge \mathcal{A}_{X}(s_1)}{(s-s_1)(s_1-s')} + \frac{s-s'}{(2 \pi i)^2} \int \frac{ds_1 \wedge \mathcal{A}_{X}(s_1)}{(s-s_1)} \int \frac{ds_2 \wedge \mathcal{A}_{X}(s_2)}{(s_1-s_2)(s_2-s')} + \ldots\,. \notag
\end{align}
Finally, note that given two distinct points $\mathcal{Z}$ and $\mathcal{Z}'$ in $\mathbb{CP}^{3|4}$ we note that the line $X$ containing them both is uniquely defined and we write
\be
U_{\mathcal{Z},\mathcal{Z}'} \equiv U\bigl(\mathcal{Z}_X(s),\mathcal{Z}_X(s')\bigr) = U_X(s,s').
\ee

To define the supertwistor version of piecewise light-like loop operators $\mathcal{L}(C)$ we introduce a sequence of supertwistors $\mathcal{Z}_i \in \mathbb{CP}^{3|4}$, for $i=1,\ldots,n$, and the corresponding sequence of intersecting lines $X_i = (\mathcal{Z}_{i-1} , \mathcal{Z}_i)$. Each line is the image of an embedding $\mathcal{Z}_{X_i}$ of $\Sigma_i = \mathbb{CP}^1$ into $\mathbb{CP}^{3|4}$ as described above,
\begin{align}
\mathcal{Z}_{X_i} : s \mapsto s \mathcal{Z}_{i-1} + \mathcal{Z}_i\,.
\end{align}
In principle, and as described in \cite{Bullimore:2011ni}, the formalism can be applied to more general curves $\Sigma_i$, but here we only consider the case where they are given by copies of $\mathbb{CP}^1$.

We may then define the holonomy around the full curve $C$ by picking a base point on one of the components ($\mathcal{Z} = s \mathcal{Z}_1 + \mathcal{Z}_n$ on the line $X_1$ for example) and writing
\begin{align}
{\rm Hol}_{\mathcal{Z}}[C] &= U_{X_1}(s,\infty)U_{X_n}(0,\infty)\ldots U_{X_2}(0,\infty) U_{X_1}(0,s) = U_{\mathcal{Z},\mathcal{Z}_n} U_{\mathcal{Z}_n,\mathcal{Z}_{n-1}}\ldots U_{\mathcal{Z}_2,\mathcal{Z}_1} U_{\mathcal{Z}_1 ,\mathcal{Z}}
\end{align}
The trace of the holonomy in some representation $R$ is then gauge-invariant and independent of the choice of $\mathcal{Z}$. Finally, the Wilson loop operator is then defined via
\be
\mathcal{L} (C) = \frac{1}{\tr_R 1\!\!1} \tr_R {\rm Hol}_{\mathcal{Z}}[C] =  \frac{1}{N} \tr \bigl[ U_{\mathcal{Z}_1 ,\mathcal{Z}_n} U_{\mathcal{Z}_n,\mathcal{Z}_{n-1}}\ldots U_{\mathcal{Z}_2,\mathcal{Z}_1} \bigr]\,,
\ee
where we have specified $R$ to be the fundamental representation for $G=\textrm{SU}(N)$ or $G=\textrm{U}(N)$.

For perturbative calculations, we make use of (\ref{UXss'}) to write,
\begin{align}
U_{\mathcal{Z}_i,\mathcal{Z}_{i-1}} = 1 - \frac{1}{2 \pi i} \int  \frac{ds_1 \wedge \mathcal{A}_{X_i}(s_1)}{s_1} - \frac{1}{(2 \pi i)^2} \int  \frac{ds_1 \wedge \mathcal{A}_{X_i}(s_1)}{s_1} \int  \frac{ds_2 \wedge \mathcal{A}_{X_i}(s_2)}{(s_1-s_2)} + \ldots\,. \notag
\end{align}

\section{Holomorphic linking and recursion relations}
\label{hololinkintro}

We now review the approach of \cite{Bullimore:2011ni} to derive the all-loop BCFW recursion of \cite{Arkani-Hamed:2010zjl} from the Wilson loop side of the duality. We begin with a family of curves as described in the previous section. While one can consider a very general family with multiple parameters, we will focus on a family of curves with a single complex parameter $t$ here. An important feature is that the dependence of the $\mathcal{Z}_{i}$ on $t$ should be holomorphic. 

Since we now have a family of curves, the component lines $X_i$ generically become dependent on $t$. We will denote the shifted line by $X_{i,t}$. As $t$ varies any such line $X_t$ will sweep out a ruled surface $\tilde{\Sigma}$ of complex dimension two, parametrised by $t$ and $s$, the coordinate for the line $\Sigma$,
\begin{align}
&f: \tilde{\Sigma} \rightarrow \mathbb{CP}^{3|4} \notag \\
&f: (t,s) \mapsto \mathcal{Z}_{X_t}(s)\,.
\end{align}

The equation (\ref{Ueq}) for the frame on the line $X_t$ at fixed $t$ is
\be
(\bar{\partial} + \mathcal{A})_{X_t} U_{X_t}(s,s') = 0\,.
\label{UeqXt}
\ee
We can now consider a variation with respect to $\bar{t}$. Writing $\bar{\delta} = d\bar{t} \,\partial_{\bar{t}}$ we have
\be
\bar{\delta} \bigl[U_{X_t}(s_1,s)(\bar{\partial} + \mathcal{A}_{X_t})U_{X_t}(s,s_0)\bigr] = 0\,.
\label{deltabarUAU}
\ee
We regard the LHS as a $(0,2)$-form living on $\tilde{\Sigma}$.
It follows that, after integrating over $s$ against $G_{s_0}(s_1,s)$ we obtain
\begin{align}
\int_\Sigma G_{s_0}(s_1,s) \wedge \Bigl\{ 
& U_{X_t}(s_1,s)(\bar{\delta}\mathcal{A}_{X_t}) U_{X_t}(s,s_0)  - U_{X_t}(s_1,s)(\bar{\partial} + \mathcal{A}_{X_t}) \wedge \bar{\delta} U_{X_t}(s,s_0)  \Bigr\} =0\,. \label{deltabarid}
\end{align}
The second term may be rewritten using 
\be
U_{X_t}(s_1,s) \mathcal{A}_{X_t} = \bar{\partial} U_{X_t}(s_1,s)\,
\label{UA=dbarU}
\ee
and integration by parts to obtain
\begin{align}
    \int_\Sigma G_{s_0}(s_1,s) \wedge U_{X_t}(s_1,s)(\bar{\delta}\mathcal{A}_{X_t}) U_{X_t}(s,s_0) &= \int_\Sigma G_{s_0}(s_1,s) \wedge \bar{\partial} \Bigl\{ U_{X_t}(s_1,s) \,\bar{\delta} U_{X_t}(s,s_0) \Bigr\} \notag \\
    &= \int_\Sigma \bar{\partial} G_{s_0}(s_1,s) \wedge \Bigl\{ U_{X_t}(s_1,s) \,\bar{\delta} U_{X_t}(s,s_0) \Bigr\} \notag \\
    &=  \bar{\delta} U_{X_t}(s_1,s_0) \,.
    \label{deltabarU}
\end{align}
Here, the final relation holds due to 
\be
\bar{\partial} G_{s_0}(s_1,s) = -i ds \wedge d \bar{s} \bigl(\delta(s-s_1) - \delta(s-s_0)\bigr) = d^2 s \bigl(\delta(s-s_1) - \delta(s-s_0)\bigr)\,.
\ee

Now, the $\bar{\delta}$-operator acting on the pullback of the connection is given by
\begin{align}
\bar{\delta} \mathcal{A}_{X_t}(s) 
&= d \bar{t} \wedge d \bar{s} \biggl[\frac{\partial \bar{Z}^{\bar{A}}}{\partial \bar{t}} \frac{\partial \bar{Z}^{\bar{B}}}{\partial \bar{s}} \partial_{\bar A} \mathcal{A}_{\bar B} + \frac{\partial^2 \bar{Z}^{\bar{A}}}{\partial \bar{t} \partial {\bar{s}}} \mathcal{A}_{\bar{A}}\biggr] \notag \\
&= \bar{\delta} \bar{Z}^{\bar A} \wedge \bar{\partial} \bar{Z}^{\bar B} \partial_{\bar A} \mathcal{A}_{\bar B}  + \bar{\delta} \bar{\partial} \bar{Z}^{\bar A}  \mathcal{A}_{\bar A} \notag \\
&= \bar{\delta} \bar{Z}^{\bar A} \wedge \bar{\partial} \bar{Z}^{\bar B} \partial_{\bar A} \mathcal{A}_{\bar B}  - \bar{\partial} \bigl[\bar{\delta} \bar{Z}^{\bar A}  \mathcal{A}_{\bar A}] - \bar{\delta} \bar{Z}^{\bar A}  \bar{\partial} \mathcal{A}_{\bar A} \notag \\
&= \bar{\delta} \bar{Z}^{\bar A} \wedge \bar{\partial} \bar{Z}^{\bar B} \bigl[\partial_{\bar A} \mathcal{A}_{\bar B}  - \partial_{\bar B} \mathcal{A}_{\bar A} \bigr] - \bar{\partial} \bigl[\bar{\delta} \bar{Z}^{\bar A}  \mathcal{A}_{\bar A}\bigr] \,. 
\label{deltabarA}
\end{align}
The first line in (\ref{deltabarA}) above is included to make clear that all terms on the RHS are pulled back to $\Sigma$ under $\mathcal{Z}_{X_t}$ so that $\mathcal{Z}$ and thus its bosonic component $Z$ depend holomorphically on both $s$, the parameter of the line and $t$, the parameter of the deformation (and hence $\bar{Z}$ depends on both anti-holomorphically).
Using this relation, equation (\ref{deltabarU}) becomes
\begin{align}
\bar{\delta} U_{X_t}(s_1,s_0) &= \int_\Sigma G_{s_0}(s_1,s) \wedge U_{X_t}(s_1,s) \bar{\delta} \bar{Z}^{\bar A} \wedge \bar{\partial} \bar{Z}^{\bar B} \bigl[\partial_{\bar A} \mathcal{A}_{\bar B} - \partial_{\bar B} \mathcal{A}_{\bar A}\bigr] U_{X_t}(s,s_0) \notag \\
&- \int_\Sigma \bar{\partial} G_{s_0}(s_1,s) U_{X_t}(s_1,s) (\bar{\delta} \bar{Z}^{\bar{A}}\mathcal{A}_{\bar{A}}) U_{X_t}(s,s_0) \notag \\
&+ \int_\Sigma  G_{s_0}(s_1,s) \bar{\partial} U_{X_t}(s_1,s) (\bar{\delta} \bar{Z}^{\bar{A}}\mathcal{A}_{\bar{A}}) U_{X_t}(s,s_0) \notag \\
&- \int_\Sigma  G_{s_0}(s_1,s)  U_{X_t}(s_1,s) (\bar{\delta} \bar{Z}^{\bar{A}}\mathcal{A}_{\bar{A}}) \bar{\partial} U_{X_t}(s,s_0) 
\end{align}
and hence we find
\begin{align}
    \overline{\mathcal{D}} U_{X_t}(s_1,s_0) &\equiv \bar{\delta} U_{X_t}(s_1,s_0) + \bar{\delta} \bar{Z}^{\bar{A}}(s_1)\mathcal{A}_{\bar{A}}\bigl(\mathcal{Z}(s_1)\bigr) U_{X_t}(s_1,s_0) - U_{X_t}(s_1,s_0) \bar{\delta} \bar{Z}^{\bar{A}}(s_0)\mathcal{A}_{\bar{A}}\bigl(\mathcal{Z}(s_0)\bigr) \notag \\
    & = \int_\Sigma G_{s_0}(s_1,s) \wedge U_{X_t}(s_1,s) \bar{\delta} \bar{Z}^{\bar A} \wedge \bar{\partial} \bar{Z}^{\bar B} \mathcal{F}_{\bar{A}\bar{B}} U_{X_t}(s,s_0)\,,
\end{align}
where 
\be
\mathcal{F}_{\bar{A} \bar{B}} = \partial_{\bar A} \mathcal{A}_{\bar B} - \partial_{\bar B} \mathcal{A}_{\bar A} + \mathcal{A}_{\bar{A}} \mathcal{A}_{\bar{B}} - \mathcal{A}_{\bar{B}} \mathcal{A}_{\bar{A}}
\ee
are the components of the $(0,2)$-form curvature,
\be
\mathcal{F} = \bar{\partial} \mathcal{A} + \mathcal{A} \wedge \mathcal{A} = \frac{1}{2} d\bar{Z}^{\bar{A}}\wedge d\bar{Z}^{\bar{B}} \mathcal{F}_{\bar{A} \bar{B}}\,.
\ee
Note that $\bar{\delta} \bar{Z}^{\bar A} \wedge \bar{\partial} \bar{Z}^{\bar B} \mathcal{F}_{\bar{A} \bar{B}}(\mathcal{Z}) $ is the pullback of the curvature $\mathcal{F}$ to $\tilde{\Sigma}$,
\be
\bar{\delta} \bar{Z}^{\bar A} \wedge \bar{\partial} \bar{Z}^{\bar B} \mathcal{F}_{\bar{A} \bar{B}}(\mathcal{Z})  = f^* \mathcal{F}(\mathcal{Z})\,, \qquad  \mathcal{Z} \equiv \mathcal{Z}_{X_{i,t}}(s)  \,.
\ee

Finally, for the full curve $C_t$, made of many components $X_{i,t}$, each of which may now vary with the parameter $t$, we have
\begin{align}
\bar{\delta} \mathcal{L} (C_t) &= \frac{1}{N} \sum_i \tr \bigl[ (\overline{\mathcal{D}}U_{X_{i,t}}) U_{X_{i-1,t}} \ldots U_{X_{i+1,t}}\bigr] \notag \\
&= \frac{1}{N} \frac{1}{2\pi i} \sum_i \int_{\Sigma_i} \frac{ds}{s} \wedge f^* \tr \bigl[ \mathcal{F}(\mathcal{Z}) {\rm Hol}_{\mathcal{Z}}[C_t]\bigr] \,,
\end{align}
where $\mathcal{Z}$ is shorthand for $\mathcal{Z}_{X_{i,t}}(s)$, i.e. the supertwistor corresponding to the image of $s$ on the line $X_{i,t}$. Note also that 
\be
G_{\infty}(0,s) = \frac{1}{2\pi i}\frac{ds}{s}\,.
\ee

We may write the expectation value of a loop operator on a curve $C$ as a path integral
\be
\langle \mathcal{L}(C)\rangle = \int D \mathcal{A} \,e^{-S_1[\mathcal{A}]}\, \mathcal{L}(C) \,.
\ee
The $\bar{\delta}$ variation for our family of curves is therefore 
\be
\bar{\delta} \langle \mathcal{L}(C_t)\rangle = \frac{1}{N} \frac{1}{2\pi i} \int D \mathcal{A} \,e^{-S_1[\mathcal{A}]} \sum_i \int_{\Sigma_i} \frac{ds}{s} \wedge  f^* \tr \bigl[ \mathcal{F}(\mathcal{Z}) {\rm Hol}_{\mathcal{Z}}[C_t]\bigr] \,.
\label{Finsertion}
\ee
Recall that if we vary the Chern-Simons action with respect to the gauge connection we find
\be
\delta S_1[\mathcal{A}] =  \beta_1 \int D^{3|4}\mathcal{Z} \wedge \tr (\mathcal{F}\wedge \delta \mathcal{A})\,.
\label{CSvariation}
\ee
Let us introduce a variational derivative $\delta/\delta \mathcal{A}$ such that
\be
\Bigl(\frac{\delta}{\delta \mathcal{A}(\mathcal{Z})}\Bigr)^{i}_{j} \Bigl(\mathcal{A} (\mathcal{Z}')\Bigr)^{k}_{l} = \Bigl(\delta_j^k \delta^i_l - \frac{\alpha}{N} \delta^i_j \delta^k_l\Bigr) \bar{\delta}^{3|4}(\mathcal{Z},\mathcal{Z}')\,.
\label{funvarA}
\ee
Here $\alpha = 0$ for a U$(N)$ gauge group while $\alpha=1$ for SU$(N)$. The object $\bar{\delta}^{3|4}(\mathcal{Z},\mathcal{Z}')$ is a distribution-valued $(0,3)$-form on the product $\mathbb{CP}^{3|4}_{\mathcal{Z}} \times \mathbb{CP}^{3|4}_{\mathcal{Z}'}$ which obeys
\be
\int_{\mathbb{CP}^{3|4}_{\mathcal{Z}'}} D^{3|4}\mathcal{Z}' \wedge \bar{\delta}^{3|4}(\mathcal{Z},\mathcal{Z}') \wedge \omega(\mathcal{Z}') = \omega(\mathcal{Z})\,
\ee
for any $(0,p)$-form $\omega$. Explicitly, we can write $\bar{\delta}^{3|4}(\mathcal{Z},\mathcal{Z}')$ as
\be
\bar{\delta}^{3|4}(\mathcal{Z},\mathcal{Z}') = \frac{1}{(2\pi)^4} \int \frac{du}{u} \wedge \bar{\delta}^{4|4}(\mathcal{Z}-u\mathcal{Z}') = \frac{1}{(2\pi)^4} \int \frac{du}{u} \bigwedge_A \bar{\partial}\frac{1}{Z^A-u Z^{\prime A}} \prod_{A'} (\chi^{A'}-u\chi^{\prime A'})\,.
\ee

From (\ref{CSvariation}) we can then write the curvature $\mathcal{F}$ as the variation of the Chern-Simons action with respect to $\mathcal{A}$,
\be
\frac{\delta}{\delta \mathcal{A}(\mathcal{Z})} S_1 [\mathcal{A}] =  \beta_1 \int D^{3|4}\mathcal{Z}' \wedge \mathcal{F}(\mathcal{Z}') \wedge \bar{\delta}^{3|4}(\mathcal{Z},\mathcal{Z}') =  \beta_1 \mathcal{F}(\mathcal{Z})\,.
\ee
and hence
\be
\bar{\delta} \langle \mathcal{L}(C_t)\rangle = -\frac{1}{\beta_1 N (2 \pi i)} \int D \mathcal{A}  \sum_i \int_{\Sigma_i} \frac{ds}{s} \wedge  f^* \tr \biggl[  \frac{\delta}{\delta \mathcal{A}(\mathcal{Z})} e^{-S_1[\mathcal{A}]} {\rm Hol}_{\mathcal{Z}}[C_t]\biggr] \,.
\label{FVariation}
\ee
As in \cite{Makeenko:1979pb,Bullimore:2011ni} we formally manipulate the path integral, using integration by parts to bring the variation to act on the holonomy,
\be
\bar{\delta} \langle \mathcal{L}(C_t)\rangle = \frac{1}{\beta_1 N (2 \pi i)} \int D \mathcal{A} \, e^{-S_1[\mathcal{A}]}\, \sum_i \int_{\Sigma_i} \frac{ds}{s} \wedge  f^*  \tr \biggl[   \frac{\delta}{\delta \mathcal{A}(\mathcal{Z})}{\rm Hol}_{\mathcal{Z}}[C_t]\biggr] \,.
\ee
To compute the action of the variational derivative on the holonomy, we may follow a similar argument as in eq. (\ref{deltabarUAU}) and below. We consider a pair of lines $X_t, X^\prime_t$, each embedded in its own copy of $\mathbb{CP}^{3|4}$. As $t$ varies, the two lines sweep out a three-dimensional surface inside $\mathbb{CP}^{3|4} \times \mathbb{CP}^{3|4}$, parametrised by $t,s,s'$. We define a map $g$ as follows
\be
g: (t,s,s') \mapsto \bigl(\mathcal{Z}_{X_t}(s),\mathcal{Z}_{X^\prime_t}(s')\bigr)\,.
\ee
Now we have (writing $\mathcal{Z} \equiv \mathcal{Z}_{X_t}(s)$ and $\mathcal{Z}' \equiv \mathcal{Z}_{X^\prime_t}(s')$)
\be
\Bigl(\frac{\delta}{\delta \mathcal{A}(\mathcal{Z})}\Bigr)^i_j  \Bigl[U_{X^\prime_t}(s_1,s')\bigl(\bar{\partial} + \mathcal{A}_{X^\prime_t}(\mathcal{Z}')\bigr)U_{X^\prime_t}(s',s_0)\Bigr]^k_l = 0\,,
\ee
where we should regard the LHS as a $(0,3)$-form on the three-dimensional space obtained by taking the direct product $\Sigma \times \Sigma'$ (i.e. the direct product of two lines) and varying $t$. 

Now, integrating over $s'$ against $G_{s_0}(s_1,s')$ we have
\begin{align}
    &\int_\Sigma G_{s_0}(s_1,s') \wedge U_{X^\prime_t}(s_1,s')^k_m \Bigl(\frac{\delta}{\delta \mathcal{A}(\mathcal{Z})}\Bigr)^i_j \mathcal{A}(\mathcal{Z}')^m_n U_{X^\prime_t}(s',s_0)^n_l \notag \\
    +&\int_\Sigma G_{s_0}(s_1,s') \wedge U_{X^\prime_t}(s_1,s')^k_m  \bigl(\bar{\partial} + \mathcal{A}_{X^\prime_t}\bigr)^m_n \Bigl(\frac{\delta}{\delta \mathcal{A}(\mathcal{Z})}\Bigr)^i_jU_{X^\prime_t}(s',s_0)^n_l = 0\,.
\end{align}
Applying (\ref{funvarA}) to the first term and (\ref{UA=dbarU}) and integration by parts to the second, we obtain
\begin{align}
&\Bigl(\frac{\delta}{\delta \mathcal{A}(\mathcal{Z})}\Bigr)^i_jU_{X}(s_1,s_0)^k_l = \notag\\ 
&-\int_\Sigma G_{s_0}(s_1,s') \wedge g^* \bar{\delta}^{3|4}(\mathcal{Z},\mathcal{Z'}) \Bigl\{U_{X}(s_1,s')^k_j  U_{X}(s',s_0)^i_l -\frac{\alpha}{N}\delta^i_j U_{X}(s_1,s_0)^k_l \Bigr\}\,. 
\label{d/dAU}
\end{align}
We can now compute the action of the variational derivative on the holonomy with the base point $\mathcal{Z}$ taken on component $i$ as follows,
\begin{align}
&\tr \biggl[   \frac{\delta}{\delta \mathcal{A}(\mathcal{Z})}{\rm Hol}_{\mathcal{Z}}[C]\biggr] = \Bigl(\frac{\delta}{\delta \mathcal{A}(\mathcal{Z})}\Bigr)^m_n \bigl[U_{X_i}(s,\infty) U_{X_{i-1}} \ldots U_{X_{i+1}} U_{X_i}(0,s)\bigr]^n_m\, \notag \\
&\quad = \Bigl(\frac{\delta}{\delta \mathcal{A}(\mathcal{Z})}\Bigr)^m_n U_{X_i}(s,\infty)^n_l \bigl[U_{X_{i-1}} \ldots U_{X_{i+1}} U_{X_i}(0,s)\bigr]^l_m \notag \\
&\quad +\sum_{j \neq i} \bigl[ U_{X_i}(s,\infty) U_{X_{i-1}}\ldots U_{X_{j+1}}\bigr]^n_r \Bigl(\frac{\delta}{\delta \mathcal{A}(\mathcal{Z})}\Bigr)^m_n \bigl(U_{X_j}\bigr)^r_k \bigl[U_{X_{j-1}} \ldots U_{X_{i+1}} U_{X_i}(0,s)\bigr]^k_m \notag \\
&\quad + \bigl[U_{X_i}(s,\infty) U_{X_{i-1}} \ldots U_{X_{i+1}}\bigr]^n_r \Bigl(\frac{\delta}{\delta \mathcal{A}(\mathcal{Z})}\Bigr)^m_n U_{X_i}(0,s)^r_m\,.
\end{align}
Making use of (\ref{d/dAU}) we find (writing $U_{X_{l,t}} = U_{X_l}$)
\begin{align}
\label{ddAHol}
&\tr \biggl[   \frac{\delta}{\delta \mathcal{A}(\mathcal{Z})}{\rm Hol}_{\mathcal{Z}}[C_t]\biggr] =  \\
& -\int_{\Sigma_i} G_{\infty}(s,s') \wedge g^* \bar{\delta}^{3|4}(\mathcal{Z},\mathcal{Z'}) \Bigl\{U_{X_i}(s,s')^n_n  U_{X_i}(s',\infty)^m_l -\frac{\alpha}{N} U_{X_i}(s,\infty)^m_l \Bigr\}\bigl[U_{X_{i-1}} \ldots U_{X_{i+1}} U_{X_i}(0,s)\bigr]^l_m \notag \\
& -\sum_{j \neq i}\int_{\Sigma_j} G_{\infty}(0,s') \wedge g^* \bar{\delta}^{3|4}(\mathcal{Z},\mathcal{Z'}) \bigl[ U_{X_i}(s,\infty) U_{X_{i-1}}\ldots U_{X_{j+1}}\bigr]^n_r  \notag \\
&\qquad \qquad  \times \Bigl\{U_{X_j}(0,s')^r_n  U_{X_j}(s',\infty)^m_k -\frac{\alpha}{N} \delta^{m}_{n} U_{X_j}(0,\infty)^r_k \Bigr\}\bigl[U_{X_{j-1}} \ldots U_{X_{i+1}} U_{X_i}(0,s)\bigr]^k_m \notag \\
& -\int_{\Sigma_i} G_{s}(0,s') \wedge g^* \bar{\delta}^{3|4}(\mathcal{Z},\mathcal{Z'}) \bigl[U_{X_i}(s,\infty) U_{X_{i-1}} \ldots U_{X_{i+1}}\bigr]^n_r \Bigl\{U_{X_i}(0,s')^r_n  U_{X_i}(s',s)^m_m -\frac{\alpha}{N} U_{X_i}(0,s)^r_n \Bigr\} \notag\,,
\end{align}
or, more compactly,
\begin{align}
&\tr \biggl[   \frac{\delta}{\delta \mathcal{A}(\mathcal{Z})}{\rm Hol}_{\mathcal{Z}}[C_t]\biggr] = \\
&-\sum_{j=1}^n \int_{\Sigma_j} \frac{ds'}{2 \pi i s'} \wedge g^*\bar{\delta}^{3|4}(\mathcal{Z},\mathcal{Z'})  
\Bigl\{\tr \bigl[U_{X_i} U_{X_{i-1}} \ldots U_{X_j} \bigr] \tr \bigl[U_{X_j} \ldots U_{X_{i+1}}U_{X_i}] -\frac{\alpha}{N} \tr \bigl[U_{X_n} \ldots U_{X_1}\bigr] \Bigr\}\,. \notag
\end{align}
Note that the first and last term of (\ref{ddAHol}) combine to give the term $j=i$ in the sum.

We therefore obtain for the variation of the expectation value of the loop operator,
\be
\bar{\delta} \langle \mathcal{L}(C_t)\rangle = -\frac{N}{\beta_1 (2 \pi i)^2} \sum_{i,j} \int_{\Sigma_i \times \Sigma_j} \frac{ds}{s} \wedge\frac{ds'}{s'} \wedge g^* \bar{\delta}^{3|4}(\mathcal{Z},\mathcal{Z'})  \Bigl[ \langle \mathcal{L}(C^\prime_t) \mathcal{L}(C^{\prime \prime}_t) \rangle - \frac{\alpha}{N^2} \langle \mathcal{L}(C_t)\rangle\Bigr]\,.
\label{loopeqs}
\ee
Here, the factor of $\bar{\delta}^{3|4}(\mathcal{Z},\mathcal{Z}')$ ensures that $\mathcal{Z}$ on the line $X_i$ and $\mathcal{Z}'$ on the line $X_j$ actually coincide. Then the loop $C^{\prime}_t$ begins and ends at the intersection point, passing through the lines $X_{i-1},\ldots X_{j+1}$ while the loop $C^{\prime \prime}_t$ begins and ends at the same intersection point and passes through the lines $X_{j-1},\ldots,X_{i+1}$.

Taking the large $N$ limit of (\ref{loopeqs}) we identify $\beta_1 \rightarrow -\frac{iC_F}{4\pi^3}$ and one obtains
\be
\bar{\delta} \langle \mathcal{L}(C_t)\rangle = 2 \pi i \sum_{i,j} \int_{\Sigma_i \times \Sigma_j} \frac{ds}{s} \wedge\frac{ds'}{s'} \wedge g^* \bar{\delta}^{3|4}(\mathcal{Z},\mathcal{Z'}) \langle \mathcal{L}(C^\prime_t) \rangle \langle \mathcal{L}(C^{\prime \prime}_t)\rangle\,.
\label{largeNloopeq}
\ee

To obtain the explicit form of the BCFW recursion relation from (\ref{largeNloopeq}) we choose an explicit deformation of the form
\be
\mathcal{Z}_n \rightarrow \hat{\mathcal{Z}}_n = \mathcal{Z}_n + t \mathcal{Z}_{n-1}\,.
\label{BCFWshift}
\ee
In this case the only line for which the pullback $f^* \mathcal{F}$ of the Chern-Simons curvature is non-vanishing is $X_{1,t}$. Therefore, the only term in the sum in (\ref{Finsertion}) is $i=1$. The map $\mathcal{Z}_{X_{1,t}}$ is given explicitly by 
\be
\mathcal{Z}_{X_{1,t}}(s) = s \mathcal{Z}_1 + \mathcal{Z}_n + t \mathcal{Z}_{n-1} \equiv \mathcal{Z}\,.
\ee
Then we integrate both sides of (\ref{largeNloopeq}) as follows,
\be
\int \frac{dt}{t} \wedge \bar{\delta} \langle \mathcal{L}(C_t)\rangle = 2 \pi i  \int \frac{dt}{t} \wedge \frac{ds}{s} \wedge \frac{ds'}{s'} \wedge \frac{du}{u} \wedge  \bar{\delta}^{4|4}(\mathcal{Z} - u\mathcal{Z'}) \langle \mathcal{L}(C^\prime_t) \rangle \langle \mathcal{L}(C^{\prime \prime}_t)\rangle\,,
\label{RecRel}
\ee
where the delta function is given explicitly by
\be
\bar{\delta}^{4|4}(\mathcal{Z} - u\mathcal{Z'})=\bar{\delta}^{4|4}\bigl(s \mathcal{Z}_1 + \mathcal{Z}_n + t \mathcal{Z}_{n-1} - u(s' \mathcal{Z}_j + \mathcal{Z}_{j-1})\bigr)\,.
\ee
The LHS of (\ref{RecRel}) is
\be
\int \frac{dt}{t} \wedge \bar{\delta} \langle \mathcal{L}(C_t)\rangle = 2 \pi i \bigl(\langle \mathcal{L}(C_{t=0})\rangle - \langle \mathcal{L}(C_{t=\infty})\rangle\bigr)
\ee
while the RHS is 
\be
2 \pi i \sum_j 
[n,n-1,1,j,j-1]
\langle \mathcal{L}(C^\prime_{t=t_j}) \rangle \langle \mathcal{L}(C^{\prime \prime}_{t=t_j})\rangle\,.
\ee
We conclude
\be
\langle \mathcal{L}[1,\ldots,n] \rangle = \langle \mathcal{L}[1,\ldots,n-1]\rangle + \sum_j [n-1,n,1,j-1,j]\langle \mathcal{L}[1,\ldots,j-1,I_j]\rangle \langle \mathcal{L}[I_j,j,\ldots,\hat{n}_j]\rangle\,.
\label{BCFWlargeN}
\ee
Here, the supertwistors $\hat{\mathcal{Z}}_n(j)$ and $\mathcal{Z}_{I}(j)$ are given by
\begin{align}
    \hat{\mathcal{Z}}_n &= \mathcal{Z}_{n-1} \langle n \,1\,j-1\,j\rangle - \mathcal{Z}_n \langle n-1\,1\,j-1\,j\rangle\,, \notag \\
    \mathcal{Z}_I &= \mathcal{Z}_{j-1} \langle j\, n-1\, n \, 1\rangle - \mathcal{Z}_j \langle j-1\, n-1\, n\, 1 \rangle\,.
\end{align}
More generally, if we return to (\ref{loopeqs}) and we do not take the large $N$ limit, we obtain (again identifying $\beta_1 \rightarrow -\frac{iC_F}{4\pi^3}$)
\begin{align}
&\langle \mathcal{L}[1,\ldots,n] \rangle = \langle \mathcal{L}[1,\ldots,n-1]\rangle  \notag \\ 
&+\sum_j [n-1,n,1,j-1,j] \Bigl[\langle \mathcal{L}[1,\ldots,j-1,I_j]  \mathcal{L}[I_j,j,\ldots,\hat{n}_j]\rangle -\frac{\alpha}{N^2} \langle \mathcal{L}[1,\ldots,\hat{n}_j] \rangle \Bigr]\,. 
\label{colourexactoneWL0}
\end{align}
Writing the correlator of two Wilson loops as the sum of disconnected and connected parts, $\langle \mathcal{L}_1 \mathcal{L}_2 \rangle = \langle \mathcal{L}_1 \rangle  \langle \mathcal{L}_2 \rangle + \langle \mathcal{L}_1 \mathcal{L}_2 \rangle^{\rm c}$, gives us the relation 
\begin{align}
\langle \mathcal{L}[1,\ldots,n] \rangle =\,\,  &\langle \mathcal{L}[1,\ldots,n-1]\rangle \label{colourexactoneWL} \\ &+ \sum_j [n-1,n,1,j-1,j] \Bigl[\langle \mathcal{L}[1,\ldots,j-1,I_j] \rangle \langle \mathcal{L}[I_j,j,\ldots,\hat{n}_j]\rangle \notag \\
&\qquad\qquad\qquad +\langle \mathcal{L}[1,\ldots,j-1,I_j]  \mathcal{L}[I_j,j,\ldots,\hat{n}_j]\rangle^{\rm c} -\frac{\alpha}{N^2} \langle \mathcal{L}[1,\ldots,\hat{n}_j] \rangle \Bigr]\,. \notag
\end{align}

In contrast to the equation (\ref{BCFWlargeN}), the full relation obtained without taking the large $N$ limit is not closed. It requires knowledge about the connected part of the correlator of two light-like Wilson loops, albeit in a special kinematic limit, where the two loops share a common `intersection' twistor $I_j$.

However, as will now argue, a similar recursion relation can be obtained for such objects following very similar logic to that presented above.

\section{Recursion for tree-level correlators of multiple loop operators}
\label{2WLrecursion}

Let us consider the correlator of two loop operators $\mathcal{L}(C)$ and $\mathcal{L}(\tilde{C})$. We will denote the twistors describing the two contours as $\mathcal{Z}_1,\ldots,\mathcal{Z}_{n_1}$ and $\tilde{\mathcal{Z}}_1, \ldots, \tilde{\mathcal{Z}}_{n_2}$.
Considering the variation of the correlator of the two loop operators in some one-parameter deformation, we find in complete analogy to (\ref{Finsertion}),
\begin{align}
\label{FinsertionLL}
&\bar{\delta} \langle \mathcal{L}(C_t) \mathcal{L}(\tilde{C}_t)\rangle = \frac{1}{N} \frac{1}{2\pi i} \int D \mathcal{A} \,e^{-S_1[\mathcal{A}]} \times \\
&\,\,\,\Bigl[\sum_i \int_{\Sigma_i} \frac{ds}{s} \wedge  f^* \tr \bigl[ \mathcal{F}(\mathcal{Z}) {\rm Hol}_{\mathcal{Z}}[C_t]\bigr] \mathcal{L}(\tilde{C}_t) + \sum_i \int_{\tilde{\Sigma}_i} \frac{d\tilde{s}}{\tilde{s}} \wedge  f^* \mathcal{L}(C_t) \tr \bigl[ \mathcal{F}(\tilde{\mathcal{Z}}) {\rm Hol}_{\tilde{\mathcal{Z}}}[\tilde{C}_t]\bigr] \Bigr] \,. \notag
\end{align}
We may employ the same logic as in the case of a single loop operator to write the insertion of the curvature as the variational derivative on the action and then integrate by parts inside the path integral. We obtain (writing $\mathcal{Z}_{X_{i,t}}(s) \equiv \mathcal{Z}$)
\begin{align}
&\bar{\delta} \langle \mathcal{L}(C_t) \mathcal{L}(\tilde{C}_t)\rangle = \frac{1}{\beta_1 N (2 \pi i)} \int D \mathcal{A} \, e^{-S_1[\mathcal{A}]} \times \notag \\
\,\,\, &\sum_i \int_{\Sigma_i} \frac{ds}{s} \wedge  f^*  \biggl[\tr \Bigl[   \frac{\delta}{\delta \mathcal{A}(\mathcal{Z})}{\rm Hol}_{\mathcal{Z}}[C_t]\Bigr] \mathcal{L}(\tilde{C}_t) + \tr \Bigl[{\rm Hol}_{\mathcal{Z}}[C_t] \frac{\delta}{\delta \mathcal{A}(\mathcal{Z})} \mathcal{L}(\tilde{C}_t)\Bigr] \biggr] + (C \leftrightarrow \tilde{C})\,.
\label{deltabarLL}
\end{align}

This results in (writing $\mathcal{Z}_{X_{i,t}}(s) \equiv \mathcal{Z}$, $\mathcal{Z}_{X_{j,t}}(s') \equiv \mathcal{Z}'$ and $\mathcal{Z}_{\tilde{X}_{j,t}}(\tilde{s}) \equiv \tilde{\mathcal{Z}}$)
\begin{align}
&\bar{\delta} \langle \mathcal{L}(C_t) \mathcal{L}(\tilde{C}_t)\rangle = \notag \\
&-\frac{N}{\beta_1 (2 \pi i)^2} \sum_{i,j} \int  \frac{ds}{s} \wedge\frac{ds'}{s'} \wedge g^* \bar{\delta}^{3|4}(\mathcal{Z},\mathcal{Z'})
\Bigl[ \langle \mathcal{L}(C^\prime_t) \mathcal{L}(C^{\prime \prime}_t) \mathcal{L}(\tilde{C}_t)\rangle - \frac{\alpha}{N^2} \langle \mathcal{L}(C_t) \mathcal{L}(\tilde{C}_t)\rangle\Bigr]  \notag \\
&-\frac{1}{\beta_1 (2 \pi i)^2 N^2} \sum_{i,j} \int  \frac{ds}{s} \wedge\frac{d\tilde{s}}{\tilde{s}} \wedge g^* \bar{\delta}^{3|4}(\mathcal{Z},\tilde{\mathcal{Z}})
\Bigl[ \langle \tr ({\rm Hol}_{\mathcal{Z}}[C_t] {\rm Hol}_{\tilde{\mathcal{Z}}}[\tilde{C}_t])\rangle - \frac{\alpha}{N} \langle \tr {\rm Hol}_{\mathcal{Z}}[C_t] \tr {\rm Hol}_{\tilde{\mathcal{Z}}}[\tilde{C}_t]\rangle \Bigr] \notag \\
& + (C \leftrightarrow \tilde{C}) \,.
\end{align}

Now the LHS can be decomposed into disconnected and connected contributions (marked with a `c' superscript),
\be
\bar{\delta} \langle \mathcal{L}(C_t) \mathcal{L}(\tilde{C}_t)\rangle =  \bar{\delta} \langle \mathcal{L}(C_t) \rangle \langle \mathcal{L}(\tilde{C}_t)\rangle + \langle \mathcal{L}(C_t) \rangle \bar{\delta} \langle \mathcal{L}(\tilde{C}_t)\rangle + \bar{\delta} \langle \mathcal{L}(C_t) \mathcal{L}(\tilde{C}_t)\rangle^{\rm c}\,.
\label{LHSdecomp}
\ee
Likewise, the various terms on the RHS may be decomposed as
\begin{align}
    \langle \mathcal{L}(C^\prime_t) \mathcal{L}(C^{\prime \prime}_t) \mathcal{L}(\tilde{C}_t)\rangle = \,\,\,\,&\langle \mathcal{L}(C^\prime_t) \rangle \langle \mathcal{L}(C^{\prime \prime}_t) \rangle \langle \mathcal{L}(\tilde{C}_t)\rangle + \langle \mathcal{L}(C^\prime_t)  \mathcal{L}(C^{\prime \prime}_t) \rangle^{\rm c} \langle \mathcal{L}(\tilde{C}_t)\rangle \notag \\
    + &\langle \mathcal{L}(C^\prime_t) \rangle \langle \mathcal{L}(C^{\prime \prime}_t) \mathcal{L}(\tilde{C}_t)\rangle^{\rm c} + \langle \mathcal{L}(C^{\prime \prime}_t) \rangle \langle \mathcal{L}(C^{\prime}_t) \mathcal{L}(\tilde{C}_t)\rangle^{\rm c} \notag \\
    + &\langle \mathcal{L}(C^\prime_t)  \mathcal{L}(C^{\prime \prime}_t) \mathcal{L}(\tilde{C}_t)\rangle^{\rm c} \\
    \langle \mathcal{L}(C_t) \mathcal{L}(\tilde{C}_t)\rangle =\,\,\,\, &\langle \mathcal{L}(C_t) \rangle \langle \mathcal{L}(\tilde{C}_t)\rangle + \langle \mathcal{L}(C_t) \mathcal{L}(\tilde{C}_t)\rangle^{\rm c} \\
    \frac{\alpha}{N} \langle \tr {\rm Hol}_{\mathcal{Z}}[C_t] \tr {\rm Hol}_{\tilde{\mathcal{Z}}}[\tilde{C}_t]\rangle = \,\,\,\, & \alpha N \langle \mathcal{L}(C_t) \mathcal{L}(\tilde{C}_t)\rangle = \alpha N\bigl(\langle \mathcal{L}(C_t) \rangle \langle \mathcal{L}(\tilde{C}_t)\rangle + \langle \mathcal{L}(C_t) \mathcal{L}(\tilde{C}_t)\rangle^{\rm c}\bigr)
\end{align}
We may then make use of (\ref{loopeqs}) for the first two terms in (\ref{LHSdecomp}) and we find various terms from the RHS are cancelled. We are left with 
\begin{align}
&\bar{\delta} \langle \mathcal{L}(C_t) \mathcal{L}(\tilde{C}_t)\rangle^{\rm c} = \notag \\
&-\frac{N}{\beta_1 (2 \pi i)^2} \sum_{i,j} \int  \frac{ds}{s} \wedge\frac{ds'}{s'} \wedge g^* \bar{\delta}^{3|4}(\mathcal{Z},\mathcal{Z'})
\Bigl[ \langle \mathcal{L}(C^\prime_t) \rangle \langle \mathcal{L}(C^{\prime \prime}_t) \mathcal{L}(\tilde{C}_t)\rangle^{\rm c} + \langle \mathcal{L}(C^{\prime \prime}_t) \rangle \langle \mathcal{L}(C^{\prime}_t) \mathcal{L}(\tilde{C}_t)\rangle^{\rm c} \notag \\
&\qquad \qquad \qquad \qquad \qquad \qquad \qquad \qquad  + \langle \mathcal{L}(C^\prime_t)  \mathcal{L}(C^{\prime \prime}_t) \mathcal{L}(\tilde{C}_t)\rangle^{\rm c} - \frac{\alpha}{N^2} \langle \mathcal{L}(C_t) \mathcal{L}(\tilde{C}_t)\rangle^{\rm c} \Bigr] \notag \\
&-\frac{1}{\beta_1 (2 \pi i)^2 N^2} \sum_{i,j} \int  \frac{ds}{s} \wedge\frac{d\tilde{s}}{\tilde{s}} \wedge g^* \bar{\delta}^{3|4}(\mathcal{Z},\tilde{\mathcal{Z}})
\Bigl[ \langle \tr ({\rm Hol}_{\mathcal{Z}}[C_t] {\rm Hol}_{\tilde{\mathcal{Z}}}[\tilde{C}_t])\rangle \notag \\
&\qquad \qquad \qquad \qquad \qquad \qquad \qquad \qquad - \alpha N \bigl(\langle \mathcal{L}(C_t) \rangle \langle \mathcal{L}(\tilde{C}_t)\rangle + \langle \mathcal{L}(C_t) \mathcal{L}(\tilde{C}_t)\rangle^{\rm c}\bigr) \Bigr] \notag \\
&+ (C \leftrightarrow \tilde{C}) \,.
\label{loopeqsLLconn}
\end{align}

Now we may again take the large $N$ limit (i.e. keep leading terms for large $N$ on both sides) and we find
\begin{align}
&\bar{\delta} \langle \mathcal{L}(C_t) \mathcal{L}(\tilde{C}_t)\rangle^{\rm c} = \notag \\
&-i \pi \sum_{i,j} \int  \frac{ds}{s} \wedge\frac{ds'}{s'} \wedge g^* \bar{\delta}^{3|4}(\mathcal{Z},\mathcal{Z'})
\Bigl[ \langle \mathcal{L}(C^\prime_t) \rangle \langle \mathcal{L}(C^{\prime \prime}_t) \mathcal{L}(\tilde{C}_t)\rangle^{\rm c} + \langle \mathcal{L}(C^{\prime \prime}_t) \rangle \langle \mathcal{L}(C^{\prime}_t) \mathcal{L}(\tilde{C}_t)\rangle^{\rm c}\Bigr] \notag \\
&-\frac{i \pi}{N^3} \sum_{i,j} \int  \frac{ds}{s} \wedge\frac{d\tilde{s}}{\tilde{s}} \wedge g^* \bar{\delta}^{3|4}(\mathcal{Z},\tilde{\mathcal{Z}})
\Bigl[ \langle \tr ({\rm Hol}_{\mathcal{Z}}[C_t] {\rm Hol}_{\tilde{\mathcal{Z}}}[\tilde{C}_t])\rangle  - \alpha N \langle \mathcal{L}(C_t) \rangle \langle \mathcal{L}(\tilde{C}_t)\rangle \Bigr] \notag \\
&+ (C \leftrightarrow \tilde{C}) \,.
\end{align}

Finally, we specify the same shift as in (\ref{BCFWshift}) with only the twistor $\mathcal{Z}_{n_1}$ on the first loop being dependent on $t$. Following similar logic as before, we find the only terms in the resulting recursion relation for the connected part of the leading large $N$ correlator of two lightlike loop operators are, identifying $\beta_1 \rightarrow -\frac{iC_F}{4\pi^3}$,
\begin{align}
\label{BCFWlargeNLL}
    &\langle \mathcal{L}[1,\ldots,n_1] \mathcal{L}[\tilde{1},\ldots,\tilde{n}_2]\rangle^{\rm c} = \langle \mathcal{L}[1,\ldots,n_1-1] \mathcal{L}[\tilde{1},\ldots,\tilde{n}_2]\rangle^{\rm c}     \\
    & + \sum [n-1,n,1,j-1,j] \Bigl(\langle \mathcal{L}[1,\ldots,j-1,I_j]\rangle \langle \mathcal{L}[I_j,j,\ldots,\hat{n}_1] \mathcal{L}[\tilde{1},\ldots,\tilde{n}_2]\rangle^{\rm c}  \notag \\
    &  \qquad \qquad \qquad \qquad \qquad \quad  \,\, +\langle \mathcal{L}[I_j,j,\ldots,\hat{n}_1] \rangle \langle \mathcal{L}[1,\ldots,j-1,I_j] \mathcal{L}[\tilde{1},\ldots,\tilde{n}_2]\rangle^{\rm c} \Bigr) \notag \\
    &  +\frac{1}{N^2} \sum [n-1,n,1,\tilde{\jmath}-1,\tilde{\jmath}] \Bigl( \langle \mathcal{L}[1,\ldots,\hat{n}_1,I_{\tilde{\jmath}},\tilde{\jmath},\ldots,\tilde{\jmath}-1,I_{\tilde{\jmath}}]\rangle - \alpha \langle \mathcal{L}[1,\ldots,\hat{n}_1] \rangle \langle \mathcal{L}[\tilde{1},\ldots,\tilde{n}_2]\rangle\Bigr) \,. \notag
\end{align}

Let us comment on the nature of the terms on the RHS of (\ref{BCFWlargeNLL}). The first term is just the connected correlator of two loop operators with one twistor removed. The second and third terms are analogous to the RHS of the recursion relation (\ref{BCFWlargeN}) and arise from locations where the deformed loop self-intersects. The fourth term is a new type of contribution which arises when the deformed loop intersects the other loop, while the fifth term accompanies the fourth in the case of SU$(N)$ gauge group. In contrast to the large $N$ recursion (\ref{BCFWlargeN}) for a single Wilson loop, the recursion relation (\ref{BCFWlargeNLL}) explicitly distinguishes between the gauge groups U$(N)$ ($\alpha=0$) and SU$(N)$ ($\alpha=1$).

\subsection{Tree-level recursion in \texorpdfstring{$\bar{Q}$}{Qbar}-equation}

The relations derived above and in section \ref{hololinkintro} prove to be very useful in performing the integration on the RHS of the $\bar{Q}$-equation for planar Wilson loop correlators and the colour exact Wilson loop \cite{Drummond:2026lvq}. The $\bar{Q}$-equation is a non-perturbative equation for Wilson loops \cite{Caron-Huot:2011dec},
\begin{equation}
    \bar{Q}^{A'}_{A}\mathcal{R}_n^{(k)}=\frac{1}{4} \Gamma_{\textrm{cusp}} \textrm{Res}_{\epsilon=0}\int [d^{2|3}\mathcal{Z}_{n+1}]^{A'}_{A}[\mathcal{R}_{n+1}^{(k+1)}-\mathcal{R}_{n}^{(k)}\mathcal{R}_{n+1}^{(1,0)}]+\textrm{cyc},
\end{equation}
where $\mathcal{R}_n^{(k)}$ is the conformally invariant, finite piece of the Wilson loop and 
\begin{align}
    &\mathcal{Z}_{n+1}=\mathcal{Z}_{n}-\epsilon \mathcal{Z}_{n-1}+C \epsilon \tau \mathcal{Z}_1 + C' \epsilon^2 \mathcal{Z}_2,\\
    &\textrm{Res}_{\epsilon=0} \int_{\tau=0}^{\tau=\infty}[d^{2|3}\mathcal{Z}_{n+1}]_{A}^{A'}=C (n-1 \hspace{1mm} n \hspace{1mm} 1)_A \textrm{Res}_{\epsilon=0}\int \epsilon d \epsilon \int_0^\infty d \tau [d^3 \chi_{n+1}]^{A'}.
    \label{1wlqbar}
\end{align}
Upon perturbative expansion, the equation relates loop-level Wilson loops to Wilson loops of a lower loop order and this has allowed for the calculation of two-loop and three-loop amplitudes \cite{He:2020vob, Li:2021bwg}. In \cite{Drummond:2025ulh}, the $\bar{Q}$-equation was extended to an arbitrary number of Wilson loops,
\begin{equation}
    \bar{Q}_A^{A'} \mathcal{R}_{n_1,\hdots,n_m}^{(k)}=\frac{1}{4} \Gamma_{\textrm{cusp}}\hspace{0.5mm} \textrm{Res}_{\epsilon=0}\sum_{r=1}^{m}\int [d^{2|3}Z_{n_r+1}]^{A'}_A \left[\mathcal{R}_{n_1,\hdots,n_{r+1},\hdots,n_m}^{(k+1)}+\mathcal{R}_{n_1,\hdots,n_m}^{(k)}\mathcal{R}_{n_1,\hdots,n_{r+1},\hdots,n_m}^{(1,0)}\right] + \textrm{cyc}_r,
\end{equation}
and (\ref{1wlqbar}) was verified to hold for colour-exact Wilson loops at $O(g^2)$. This verification for multiple Wilson loops and the colour-exact Wilson loop was carried out by making use of (\ref{colourexactoneWL0}) and (\ref{BCFWlargeNLL}), respectively. By using the BCFW shift\begin{equation}
    \mathcal{Z}_{n_r}\rightarrow \mathcal{Z}_{n_r}+t \mathcal{Z}_{n_r-1}
\end{equation}
on the correlators which appear in the equations above, the only non-zero terms are those for which the $\chi_{n_r+1}$ integration is isolated to the R invariant factors which appear in the BCFW expression (this generates the $\epsilon$ poles) and the correlator parts go smoothly in the $\epsilon\rightarrow 0$ limit. For the colour exact Wilson loop $\mathcal{R}_{n}^{1-\textrm{loop}}$, the RHS of the $\bar{Q}$-equation is thus given by
\begin{align}
    &\bar{Q}_A^{A'}\mathcal{R}_n^{1-\textrm{loop}}\notag\\
    &=\frac{2 C_F}{N} \sum_{i=3}^{n-2} \log \frac{\langle n\hspace{0.5mm}1 \hspace{0.5mm} i-1 \hspace{0.5mm} i\rangle}{\langle n-1 \hspace{0.5mm} n \hspace{0.5mm} i-1 \hspace{0.5mm} i\rangle }\bar{Q}\log \frac{\langle n-1 \hspace{0.5mm} n \hspace{0.5mm} 1 \hspace{0.5mm} i \rangle}{\langle n-1 \hspace{0.5mm}n \hspace{0.5mm} 1 \hspace{0.5mm} i-1 \rangle} \Big(\langle \mathcal{L}[n,1,\hdots,i-1,\bar{I}_i]\mathcal{L}[\bar{I}_i,i,\hdots,n]\rangle^{\textrm{tree}}\notag \\
    &\hspace{5mm}-\frac{\alpha}{N^2}\langle \mathcal{L}[n,1,\hdots,i-1,\bar{I}_i,i,\hdots,n,\bar{I}_i]\rangle^{\textrm{tree}}+\Big(1-\frac{1}{N^2}\Big)\langle \mathcal{L}[1,\hdots,n]\rangle^{\textrm{tree}}\Big)+\textrm{cyc},
\end{align}
where the correlators on the RHS are taken to be the conformally invariant finite pieces, $\bar{I}_i\equiv(i-1\hspace{0.5mm} i)\cap (n-1\hspace{0.5mm} n\hspace{0.5mm} 1)$ and the Wilson loops on each side of the equation include all Grassmann sectors. Full details on the integration in the $\bar{Q}$ equation are included in section 6.2 and appendix B of \cite{Drummond:2026lvq}. 

\section{Recursion for loop-level integrands}

We recall that to complete the self-dual theory to the full maximally supersymmetric Yang-Mills theory, we should add to the Chern-Simons action the non-local interactions given by 
\be
S_2[\mathcal{A}] = \beta_2  \int d^{4|8} X \log \det\bigl(\overline{\partial} + \mathcal{A} \bigr)_X = -\beta_2 \int d^{4|8}X \sum_{r=2}^{\infty}\frac{1}{r}  
\tr (-\bar{\partial}_X^{-1} \mathcal{A})^r\,.
\ee
Therefore when we perform the path integral in the full $\mathcal{N}=4$ theory we obtain an additional term when writing the insertion of the curvature as the variation of the action,
\be
 \mathcal{F}(\mathcal{Z}) = \frac{1}{\beta_1} \biggl[
\frac{\delta S [\mathcal{A}]}{\delta \mathcal{A}(\mathcal{Z})}  - \frac{\delta S_2[\mathcal{A}]}{\delta \mathcal{A}(\mathcal{Z})}  \biggr]\,.
\ee

Making a variation of $S_2$ with respect to the connection, we find
\begin{align}
\delta S_2[\mathcal{A}] &= -\beta_2 \int d^{4|8}X \sum_{r=2}^{\infty} \Bigl( \frac{1}{2\pi i} \Bigr)^r \tr \int_{\Sigma^r} \frac{ds_1 \wedge \mathcal{A}_X(s_1) \wedge \ldots \wedge ds_{r} \wedge \delta \mathcal{A}_X(s_r)}{(s_r-s_1)\ldots(s_{r-1}-s_r)} \notag \\
&= \beta_2 \int d^{4|8}X \int_\Sigma \tr J_\mathcal{A} \wedge \delta \mathcal{A}
\end{align}
where
\be
J_{\mathcal{A}} = - \sum_{r=2} \Bigl(\frac{1}{2 \pi i}\Bigr)^r ds \int_{\Sigma^{r-1}} \frac{ds_1 \wedge \mathcal{A}_X(s_1) \wedge \ldots \wedge ds_{r-1} \wedge \mathcal{A}_X(s_{r-1})}{(s-s_1)(s_1-s_2)\ldots (s_{r-2}-s_{r-1})(s_{r-1}-s)}\,.
\label{JA}
\ee

As described in \cite{Bullimore:2011ni}, based on general arguments of \cite{Quillen}, the variation of the log-det term is given by
\be
\delta \log \det (\bar{\partial}+\mathcal{A})_X = \int_\Sigma \tr (J_{\mathcal{A}} \wedge \delta \mathcal{A}),
\ee
with
\begin{align}
J_{\mathcal{A}}(s) = \lim_{s' \rightarrow s}\bigl(G_{\mathcal{A}}(s',s) - G_0(s',s)\bigr) \notag 
\end{align}
where $G_\mathcal{A}$ is the Green's function for the operator $(\bar{\partial} + \mathcal{A})_X$ and $\mathcal{G}_0(s',s)$ is the Green's function for the $\bar{\partial}$ operator given in (\ref{Greensfn}). Since the two operators are related by the holomorphic frame $H(X,s)$ via
\be
H^{-1}(X,s) (\bar{\partial}+\mathcal{A})_X H(X,s) = \bar{\partial}_X,
\ee
the Green's function $G_{\mathcal{A}}$ is given by
\be
G_\mathcal{A}(s',s) = H(X,s') G_0(s',s) H^{-1}(X,s)
\ee
as an integral operator. We therefore find
\begin{align}
J_\mathcal{A} (s) = -\frac{ds}{2\pi i} H'(X,s) H^{-1}(X,s) = -\frac{ds}{2 \pi i} \frac{\partial U_X(s',s)}{\partial s'} \bigg|_{s'=s}\,.
\end{align}
Equivalently, we may write
\begin{align}
J_\mathcal{A} (s) = -\frac{ds}{2 \pi i} \lim_{s' \rightarrow s}\frac{U_X(s',s)-U_X(s,s)}{s'-s} = - \frac{ds}{(2 \pi i)^2} \oint \frac{ds'}{(s'-s)^2} U_X(s',s) \,.
\end{align}
From (\ref{UXss'}) we find explicitly 
\begin{align}
J_{\mathcal{A}}(s) = -\frac{ds}{2 \pi i} \biggl[ \frac{1}{2 \pi i} \int  \frac{ds_1 \wedge \mathcal{A}_{X}(s_1)}{(s-s_1)(s_1-s)} + \frac{1}{(2 \pi i)^2} \int \frac{ds_1 \wedge \mathcal{A}_{X}(s_1)}{(s-s_1)} \int \frac{ds_2 \wedge \mathcal{A}_{X}(s_2)}{(s_1-s_2)(s_2-s)} + \ldots \biggr]\,, \notag
\end{align}
matching the result in (\ref{JA}) above. Note that a simple argument shows that $\tr(J_\mathcal{A})=0$ in the SU$(N)$ case.

As explained in \cite{Bullimore:2011ni}, it is helpful to express the variation using
\be
J_{\mathcal{A}} = - \frac{1}{(2\pi i)^3}\int_{S^1 \times S^1} \frac{ds \wedge ds' \wedge ds''}{(s-s')(s'-s'')(s''-s)}  U_X(s',s) 
\ee
where the integration over $s''$ is taken around the pole at $s''=s'$. This representation is useful as we may think of the line $X$ as the image of $\Sigma = \mathbb{CP}^1$ under the embedding $\mathcal{Z}_X$ such that $\mathcal{Z}_A$ and $\mathcal{Z}_B$ are the images of $s''$ and $s'$ respectively, i.e. we write 
\be
\mathcal{Z}_X(s) = \frac{s-s'}{s''-s'}\mathcal{Z}_A + \frac{s-s''}{s'-s''}\mathcal{Z}_B\,.
\ee
The integrations force both $\mathcal{Z}_A$ and $\mathcal{Z}_B$ to coincide with $\mathcal{Z}$ on the line $X$.

We may then express $J_{\mathcal{A}}$ as
\be
J_{\mathcal{A}} = \frac{1}{(2\pi i)^2} \int_{S^1 \times S^1} G_{s''}(s',s) \wedge \frac{ds' \wedge ds''}{(s'-s'')^2} U_X(s',s) \,.
\ee
The integrations over $s'$ and $s''$ are the integrations over the locations of $\mathcal{Z}_A$ and $\mathcal{Z}_B$ over the (fixed) line $X$. They combine with integration over the line $X$ as
\be
d^{4|8}X \wedge \frac{ds' \wedge ds''}{(s'-s'')^2} = -d^{4|8}X \wedge \frac{\langle A dA \rangle \wedge \langle B d B\rangle}{\langle A B \rangle^2} = -D^{3|4}\mathcal{Z}_A \wedge D^{3|4} \mathcal{Z}_B\,.
\label{lambdaABmeasure}
\ee

Returning to (\ref{FVariation}) and working now with the full action of the $\mathcal{N}=4$ theory, we have
\begin{align}
\bar{\delta} \langle \mathcal{L}(C_t)\rangle &= -\frac{1}{\beta_1 N (2 \pi i)} \int D \mathcal{A}  \sum_i \int_{\Sigma_i} \frac{ds_i}{s_i} \wedge  f^* \tr \biggl[  \frac{\delta}{\delta \mathcal{A}(\mathcal{Z})} e^{-S[\mathcal{A}]} {\rm Hol}_{\mathcal{Z}}[C_t] + \frac{\delta S_2[\mathcal{A}]}{\delta \mathcal{A}(\mathcal{Z})}  {\rm Hol}_{\mathcal{Z}}[C_t] e^{-S[\mathcal{A}]}\biggr] \notag \\
&=\frac{1}{\beta_1 N (2 \pi i)}\int D \mathcal{A}\, e^{-S[\mathcal{A}]} \sum_i \int_{\Sigma_i} \frac{ds_i}{s_i} \wedge  f^* \tr \biggl[  \frac{\delta}{\delta \mathcal{A}(\mathcal{Z})} {\rm Hol}_{\mathcal{Z}}[C_t] - \frac{\delta S_2[\mathcal{A}]}{\delta \mathcal{A}(\mathcal{Z})}  {\rm Hol}_{\mathcal{Z}}[C_t] \biggr]\,.
\end{align}
The first term above is as analysed previously while the second term is the new contribution. It can be expressed as
\be
-\frac{\beta_2}{\beta_1 N (2 \pi i)} \int d^{4|8}X \sum_i \int_{{}_{\Sigma_i \times \Sigma}} \tilde{g}^* \bar{\delta}^{3|4}(\mathcal{Z},\hat{\mathcal{Z}}) \wedge \frac{ds_i}{s_i} \wedge 
\big\langle \! \tr \bigl[ J_{\mathcal{A}}(s) {\rm Hol}_{\mathcal{Z}}[C_t] \bigr] \big\rangle\ 
\,.
\label{newloopterm}
\ee
Here $\tilde{g}$ is the map from $\tilde{\Sigma}_i \times \Sigma$ to $\mathbb{CP}^{3|4} \times \mathbb{CP}^{3|4}$ given by
\be
\tilde{g}: (t,s_i,s) \mapsto (\mathcal{Z}_{X_{i,t}}(s_i) , \mathcal{Z}_X(s)) \equiv (\mathcal{Z},\hat{\mathcal{Z}})\,.
\ee
The presence of $\bar{\delta}^{3|4}(\mathcal{Z},\hat{\mathcal{Z}})$ forces $\mathcal{Z}$ on the line $X_{i,t}$ and $\hat{\mathcal{Z}}$ on the integration line $X$ to coincide. 

Using the form of $J_{\mathcal{A}}$ given above, we find that (\ref{newloopterm}) can be expressed as
\begin{align}
\label{newloopterm2}
-\frac{\beta_2}{\beta_1 N (2 \pi i)^3} \int d^{4|8}X & \wedge \frac{ds' \wedge ds''}{(s'-s'')^2} \times \\
&\sum_i \int_{{}_{\Sigma_i \times \Sigma}}  \tilde{g}^* \bar{\delta}^{3|4}(\mathcal{Z},\hat{\mathcal{Z}}) \wedge \frac{ds_i}{s_i} \wedge G_{s''}(s',s)   
\big\langle \tr \bigl[U_X(s',s) {\rm Hol}_{\mathcal{Z}}[C_t]\bigr] \big\rangle 
\,. \notag
\end{align}

Since the delta function and the contour integration forces $\mathcal{Z} = \hat{\mathcal{Z}} = \mathcal{Z}_B$, the trace in the first term in (\ref{newloopterm2}) can then be replaced with a new loop operator
\be
\tr \bigl[ U_{\mathcal{Z}_{i-1},\mathcal{Z}_{i-2}} \ldots U_{\mathcal{Z}_{i+1},\mathcal{Z}_{i}} U_{\mathcal{Z}_{i},\mathcal{Z}_B} U_{\mathcal{Z}_B,\hat{\mathcal{Z}}}U_{\hat{\mathcal{Z}},\mathcal{Z}_{i-1}} \bigr] = N \mathcal{L}(\tilde{C}_t \cup X) \,.
\ee

Let us now specify the same shift (\ref{BCFWshift}) as before. Once again, the only contributing term in the sum over $i$ is the term where $i=1$.
We then integrate against $dt/t$ as before and find the new contribution 
\begin{align}
    \frac{N \beta_2}{\beta_1 N (2 \pi i)^4} \int d^{4|8}X \wedge \frac{ds' \wedge ds''}{(s'-s'')^2} \int &\frac{dt}{t}\wedge\frac{ds_i}{s_i} \wedge \frac{ds (s'-s'')}{(s'-s)(s-s'')} \wedge \frac{du}{u} \wedge \bar{\delta}^{4|4}(\mathcal{Z} - u \hat{\mathcal{Z}}) \langle \mathcal{L}(\tilde{C}_t \cup X) \rangle \,.
\end{align}
Here we have explicitly
\be
\bar{\delta}^{4|4}(\mathcal{Z} - u \hat{\mathcal{Z}}) = \bar{\delta}^{4|4}\biggl(s_i \mathcal{Z}_1 + \mathcal{Z}_n + t \mathcal{Z}_{n-1} - u \Bigl(\frac{s-s'}{s''-s'}\mathcal{Z}_A + \frac{s-s''}{s'-s''}\mathcal{Z}_B\Bigr)\biggr)
\ee
We may bring this to a canonical form with a few simple steps. First we rescale $u$ by $(s''-s')$ to clear the denominators in the final two terms in the argument of the $\bar{\delta}^{4|4}$. We then define $\tilde{s} = s-s'$. This yields an integral of the form
\be
-\int \frac{dt}{t} \wedge \frac{ds_i}{s_i} \wedge \frac{d \tilde{s}}{\tilde{s}} \frac{s'-s''}{\tilde{s}+s'-s''} \wedge \frac{du}{u} \, \bar{\delta}^{4|4}\bigl(s_i \mathcal{Z}_1 + \mathcal{Z}_n + t \mathcal{Z}_{n-1} - u \tilde{s} \mathcal{Z}_A + u(\tilde{s}+s'-s'')\mathcal{Z}_B\bigr)\,.
\ee
We then further rescale, defining $\check{s} = u \tilde{s}$ and then $\check{u} = u (s'-s'')$, giving
\be
-\int \frac{dt}{t} \wedge \frac{ds_i}{s_i} \wedge \frac{d \check{s}}{\check{s}} \wedge \frac{d\check{u}}{\check{s}+\check{u}} \, \bar{\delta}^{4|4}\bigl(s_i \mathcal{Z}_1 + \mathcal{Z}_n + t \mathcal{Z}_{n-1} -  \check{s} \mathcal{Z}_A + (\check{s}+\check{u})\mathcal{Z}_B\bigr)\,.
\ee
Finally defining $\hat{u} = \check{u} + \check{s}$ yields the canonical form which we recognise as the $R$-invariant,
\begin{align}
-\int \frac{dt}{t} \wedge \frac{ds_i}{s_i} \wedge \frac{d \check{s}}{\check{s}} \wedge \frac{d\hat{u}}{\hat{u}} \, \bar{\delta}^{4|4}\bigl(s_i \mathcal{Z}_1 + \mathcal{Z}_n + t \mathcal{Z}_{n-1} -  \check{s} \mathcal{Z}_A + \hat{u}\mathcal{Z}_B\bigr) &= -[n,n-1,1,A,B] \notag \\
&= [n-1,n,1,A,B]
\end{align}
Putting everything together, we obtain the following recursion relation 
\begin{align}
&\langle \mathcal{L}[1,\ldots,n] \rangle = \langle \mathcal{L}[1,\ldots,n-1]\rangle  \notag \\ 
&+\sum_j [n-1,n,1,j-1,j] \Bigl[\langle \mathcal{L}[1,\ldots,j-1,I_j]  \mathcal{L}[I_j,j,\ldots,\hat{n}_j]\rangle -\frac{\alpha}{N^2} \langle \mathcal{L}[1,\ldots,\hat{n}_j] \rangle \Bigr] \notag \\
&-\frac{\beta_2}{\beta_1  (2 \pi i)^5} \int D^{3|4}\mathcal{Z}_A \wedge D^{3|4}\mathcal{Z}_B [n-1,n,1,A,B] \langle \mathcal{L}[1,\ldots,n-1,\hat{n}_{AB},\hat{\mathcal{Z}},B] \rangle \,.
\end{align}
Taking the large $N$ limit yields 
\begin{align}
&\langle \mathcal{L}[1,\ldots,n] \rangle = \langle \mathcal{L}[1,\ldots,n-1]\rangle  + \sum_j [n-1,n,1,j-1,j] \langle \mathcal{L}[1,\ldots,j-1,I_j] \rangle \langle  \mathcal{L}[I_j,j,\ldots,\hat{n}_j]\rangle  \notag \\
&\qquad \qquad +\frac{ g^2}{\pi^2} \frac{1}{(2 \pi i)^2} \int D^{3|4}\mathcal{Z}_A \wedge D^{3|4}\mathcal{Z}_B [n-1,n,1,A,B] \langle \mathcal{L}[1,\ldots,n-1,\hat{n}_{AB},\hat{\mathcal{Z}},B] \rangle \,.
\end{align}
Note that the \(\frac{1}{(2 \pi i)^2}\) factor can be thought of as part of the residual GL(2) integral which is to be performed on the last line, which will be cancelled when computing that integral using residues. As discussed in \cite{Bullimore:2011ni}, this matches the BCFW recursion for the loop integrand obtained in \cite{Arkani-Hamed:2010zjl}.

Now, let us repeat the analysis in the case of the correlator of two Wilson loops. We return to eq. (\ref{deltabarLL}) but now find an additional term on the RHS,
\begin{align}
&\bar{\delta} \langle \mathcal{L}(C_t) \mathcal{L}(\tilde{C}_t)\rangle = \frac{1}{\beta_1 N (2 \pi i)} \int D \mathcal{A} \, e^{-S_1[\mathcal{A}]} \times \notag \\
 &\sum_i \int_{\Sigma_i} \frac{ds_i}{s_i} \wedge  f^*  \biggl[\tr \Bigl[   \frac{\delta}{\delta \mathcal{A}(\mathcal{Z})}{\rm Hol}_{\mathcal{Z}}[C_t] - \frac{\delta S_2[\mathcal{A}]}{\delta \mathcal{A}(\mathcal{Z})}  {\rm Hol}_{\mathcal{Z}}[C_t] \Bigr] \mathcal{L}(\tilde{C}_t) + \tr \Bigl[{\rm Hol}_{\mathcal{Z}}[C_t] \frac{\delta}{\delta \mathcal{A}(\mathcal{Z})} \mathcal{L}(\tilde{C}_t)\Bigr] \biggr] \notag \\
&+ (C \leftrightarrow \tilde{C})\,.
\end{align}
The new contribution can be expressed as
\begin{align}
\label{newlooptermLL}
-\frac{\beta_2}{\beta_1 N (2 \pi i)} \int d^{4|8}X \sum_i \int_{{}_{\Sigma_i \times \Sigma}} \tilde{g}^* \bar{\delta}^{3|4}(\mathcal{Z},\hat{\mathcal{Z}}) \wedge \frac{ds_i}{s_i} \wedge \big\langle \! \tr \bigl[ J_{\mathcal{A}}(s) {\rm Hol}_{\mathcal{Z}}[C_t] \bigr] \mathcal{L}(\tilde{C}_t)\big\rangle\  + (C \leftrightarrow \tilde{C}) \,.
\end{align}
As before, we can decompose all contributions into disconnected and connected parts. This results in an equation for the connected part of the correlator,
\begin{align}
&\bar{\delta} \langle \mathcal{L}(C_t) \mathcal{L}(\tilde{C}_t)\rangle^{\rm c} = \notag \\
&-\frac{N}{\beta_1 (2 \pi i)^2} \sum_{i,j} \int  \frac{ds}{s} \wedge\frac{ds'}{s'} \wedge g^* \bar{\delta}^{3|4}(\mathcal{Z},\mathcal{Z'})
\Bigl[ \langle \mathcal{L}(C^\prime_t) \rangle \langle \mathcal{L}(C^{\prime \prime}_t) \mathcal{L}(\tilde{C}_t)\rangle^{\rm c} + \langle \mathcal{L}(C^{\prime \prime}_t) \rangle \langle \mathcal{L}(C^{\prime}_t) \mathcal{L}(\tilde{C}_t)\rangle^{\rm c} \notag \\
&\qquad \qquad \qquad \qquad \qquad \qquad \qquad \qquad  + \langle \mathcal{L}(C^\prime_t)  \mathcal{L}(C^{\prime \prime}_t) \mathcal{L}(\tilde{C}_t)\rangle^{\rm c} - \frac{\alpha}{N^2} \langle \mathcal{L}(C_t) \mathcal{L}(\tilde{C}_t)\rangle^{\rm c} \Bigr] \notag \\
&-\frac{1}{\beta_1 N^2 (2 \pi i)^2} \sum_{i,j} \int  \frac{ds}{s} \wedge\frac{d\tilde{s}}{\tilde{s}} \wedge g^* \bar{\delta}^{3|4}(\mathcal{Z},\tilde{\mathcal{Z}})
\Bigl[ \langle \tr ({\rm Hol}_{\mathcal{Z}}[C_t] {\rm Hol}_{\tilde{\mathcal{Z}}}[\tilde{C}_t])\rangle \notag \\
&\qquad \qquad \qquad \qquad \qquad \qquad \qquad \qquad - \alpha N \bigl(\langle \mathcal{L}(C_t) \rangle \langle \mathcal{L}(\tilde{C}_t)\rangle + \langle \mathcal{L}(C_t) \mathcal{L}(\tilde{C}_t)\rangle^{\rm c}\bigr) \Bigr] \notag \\
&-\frac{\beta_2}{\beta_1 N (2 \pi i)} \int d^{4|8}X \sum_i \int_{{}_{\Sigma_i \times \Sigma}} \tilde{g}^* \bar{\delta}^{3|4}(\mathcal{Z},\hat{\mathcal{Z}}) \wedge \frac{ds_i}{s_i} \wedge \big\langle \! \tr \bigl[ J_{\mathcal{A}}(s) {\rm Hol}_{\mathcal{Z}}[C_t] \bigr] \mathcal{L}(\tilde{C}_t)\big\rangle^{\rm c}  \notag \\
&+ (C \leftrightarrow \tilde{C}) \,.
\end{align}
In the large $N$ limit the following terms survive
\begin{align}
&\bar{\delta} \langle \mathcal{L}(C_t) \mathcal{L}(\tilde{C}_t)\rangle^{\rm c} = \notag \\
&-\frac{N}{\beta_1 (2 \pi i)^2} \sum_{i,j} \int  \frac{ds}{s} \wedge\frac{ds'}{s'} \wedge g^* \bar{\delta}^{3|4}(\mathcal{Z},\mathcal{Z'})
\Bigl[ \langle \mathcal{L}(C^\prime_t) \rangle \langle \mathcal{L}(C^{\prime \prime}_t) \mathcal{L}(\tilde{C}_t)\rangle^{\rm c} + \langle \mathcal{L}(C^{\prime \prime}_t) \rangle \langle \mathcal{L}(C^{\prime}_t) \mathcal{L}(\tilde{C}_t)\rangle^{\rm c}  \Bigr] \notag \\
&-\frac{1}{\beta_1 N^2 (2 \pi i)^2} \sum_{i,j} \int  \frac{ds}{s} \wedge\frac{d\tilde{s}}{\tilde{s}} \wedge g^* \bar{\delta}^{3|4}(\mathcal{Z},\tilde{\mathcal{Z}})
\Bigl[ \langle \tr ({\rm Hol}_{\mathcal{Z}}[C_t] {\rm Hol}_{\tilde{\mathcal{Z}}}[\tilde{C}_t])\rangle  - \alpha N \bigl(\langle \mathcal{L}(C_t) \rangle \langle \mathcal{L}(\tilde{C}_t)\rangle  \Bigr] \notag \\
&-\frac{\beta_2}{\beta_1 N (2 \pi i)} \int d^{4|8}X \sum_i \int_{{}_{\Sigma_i \times \Sigma}} \tilde{g}^* \bar{\delta}^{3|4}(\mathcal{Z},\hat{\mathcal{Z}}) \wedge \frac{ds_i}{s_i} \wedge \big\langle \! \tr \bigl[ J_{\mathcal{A}}(s) {\rm Hol}_{\mathcal{Z}}[C_t] \bigr] \mathcal{L}(\tilde{C}_t)\big\rangle^{\rm c}  \notag \\
&+ (C \leftrightarrow \tilde{C}) \,.
\end{align}
Taking the same shift as before and then performing the integration against $dt/t$ and using the large $N$ limits for $\beta_1$ and $\beta_2$ yields the following recursion relation for the connected correlator in the large $N$ limit,
\begin{align}
    &\langle \mathcal{L}[1,\ldots,n_1] \mathcal{L}[\tilde{1},\ldots,\tilde{n}_2]\rangle^{\rm c} = \langle \mathcal{L}[1,\ldots,n_1-1] \mathcal{L}[\tilde{1},\ldots,\tilde{n}_2]\rangle^{\rm c}  \notag \\
    &\quad + \sum [n-1,n,1,j-1,j] \Bigl(\langle \mathcal{L}[1,\ldots,j-1,I_j]\rangle \langle \mathcal{L}[I_j,j,\ldots,\hat{n}_1] \mathcal{L}[\tilde{1},\ldots,\tilde{n}_2]\rangle^{\rm c}  \notag \\
    &  \qquad \qquad \qquad \qquad \qquad \quad \,\,\, +\langle \mathcal{L}[I_j,j,\ldots,\hat{n}_1] \rangle \langle \mathcal{L}[1,\ldots,j-1,I_j] \mathcal{L}[\tilde{1},\ldots,\tilde{n}_2]\rangle^{\rm c} \Bigr) \notag \\
    & \quad+\frac{1}{N^2} \sum [n-1,n,1,\tilde{\jmath}-1,\tilde{\jmath}] \Bigl( \langle \mathcal{L}[1,\ldots,\hat{n}_1,I_{\tilde{\jmath}},\tilde{\jmath},\ldots,\tilde{\jmath}-1,I_{\tilde{\jmath}}]\rangle - \alpha \langle \mathcal{L}[1,\ldots,\hat{n}_1] \rangle \langle \mathcal{L}[\tilde{1},\ldots,\tilde{n}_2]\rangle\Bigr)  \notag \\
    &\quad +\frac{g^2}{\pi^2} \frac{1}{(2 \pi i)^2} \int D^{3|4}\mathcal{Z}_A \wedge D^{3|4}\mathcal{Z}_B [n-1,n,1,A,B] \langle \mathcal{L}[1,\ldots,n-1,\hat{n}_{AB},\hat{\mathcal{Z}},B] \mathcal{L}[\tilde{1},\ldots,\tilde{n}_2]\rangle^{\rm c} \, .
    \label{looprecursion}
\end{align}

The integrand of an $l$-loop Wilson loop correlator is given by a correlator with Lagrangian insertions $\mathcal{L}(X_i)=\sum_{r=2}^\infty \frac{1}{r}(-\bar{\partial}_{X_i}^{-1}\mathcal{A})^r$,
\begin{equation}
    \langle \mathcal{L}(C_1) \hdots \mathcal{L}(C_n)\rangle^{l-\textrm{loop}}=\frac{1}{l!}\int \bigwedge_{i=1}^l d^{4|8}X_i \langle  \mathcal{L}(C_1) \hdots \mathcal{L}(C_n) \mathcal{L}(X_1) \hdots \mathcal{L}(X_l)\rangle^{\textrm{CS}},
    \label{loopintegrand}
\end{equation}
where the correlator on the RHS is computed in holomorphic Chern-Simons theory (i.e. without $S_2[\mathcal{A}]$). To shorten notation, we label $\mathcal{L}(X_i)$ by $\mathcal{L}_i$. 

As it is written, (\ref{looprecursion}) depends on loop-integrated correlators. Using (\ref{loopintegrand}), we can express it in terms of loop integrands,
\begin{align}
    &\langle \mathcal{L}[1,\ldots,n_1] \mathcal{L}[\tilde{1},\ldots,\tilde{n}_2]\mathcal{L}_1\hdots\mathcal{L}_l\rangle^{\rm c} = \langle \mathcal{L}[1,\ldots,n_1-1] \mathcal{L}[\tilde{1},\ldots,\tilde{n}_2]\mathcal{L}_1\hdots\mathcal{L}_l\rangle^{\rm c}  \notag \\
    &\quad + \sum_{j,p} [n-1,n,1,j-1,j] \Bigl(\langle \mathcal{L}[1,\ldots,j-1,I_j]\mathcal{L}_{p_L}\rangle \langle \mathcal{L}[I_j,j,\ldots,\hat{n}_1] \mathcal{L}[\tilde{1},\ldots,\tilde{n}_2]\mathcal{L}_{p_R}\rangle^{\rm c}  \notag \\
    &  \qquad \qquad \qquad \qquad \qquad \quad \,\,\, +\langle \mathcal{L}[I_j,j,\ldots,\hat{n}_1]\mathcal{L}_{p_L} \rangle \langle \mathcal{L}[1,\ldots,j-1,I_j] \mathcal{L}[\tilde{1},\ldots,\tilde{n}_2]\mathcal{L}_{p_R}\rangle^{\rm c} \Bigr) \notag \\
    & \quad +\frac{1}{N^2} \sum_{\tilde{\jmath}} [n-1,n,1,\tilde{\jmath}-1,\tilde{\jmath}] \Bigl( \langle \mathcal{L}[1,\ldots,\hat{n}_1,I_{\tilde{\jmath}},\tilde{\jmath},\ldots,\tilde{\jmath}-1,I_{\tilde{\jmath}}]\mathcal{L}_1 \hdots \mathcal{L}_l\rangle\\
    & \qquad \qquad \qquad \qquad \qquad \,\,\,\,\,- \alpha \sum_{p} \langle \mathcal{L}[1,\ldots,\hat{n}_1]\mathcal{L}_{p_L} \rangle \langle \mathcal{L}[\tilde{1},\ldots,\tilde{n}_2]\mathcal{L}_{p_L}\rangle\Bigr)  \notag \\
    &\quad +\frac{1}{(2 \pi i)^2} \sum_{l_0=1}^l\int_{\textrm{GL}(2)_{l_0}}  [n-1,n,1,A'_{l_0},B'_{l_0}] \langle \mathcal{L}[1,\ldots,n-1,\hat{n}_{A_{l_0}B_{l_0}},\hat{\mathcal{Z}},B'_{l_0}] \mathcal{L}[\tilde{1},\ldots,\tilde{n}_2]\prod_{i\in L\backslash\{l_0\}} \mathcal{L}_{i}\rangle^{\rm c} \, .
    \label{looprecursionintegrand}
\end{align}
We have suppressed the ``CS" label on each correlator. We sum over the partitions $p$ of $\{\mathcal{L}_1,\hdots,\mathcal{L}_l\}$ into two sets, $p_L$ and $p_R$, with $\mathcal{L}_{p_L}=\prod_{i\in p_L}\mathcal{L}_i$, $\mathcal{L}_{p_R}=\prod_{i\in p_R}\mathcal{L}_i$. Note that the sum is symmetric under permutations of the Langrangian insertions. As an explicit example, for $l=2$, the partitions are
\begin{equation}
    \left\{\left\{ \{\mathcal{L}_1,\mathcal{L}_2\},\{\}\right\},\left\{ \{\mathcal{L}_1\},\{\mathcal{L}_2\}\right\},\left\{ \{\mathcal{L}_2\},\{\mathcal{L}_1\}\right\},\left\{\{\},\{\mathcal{L}_1,\mathcal{L}_2\}\right\}\right\}.
\end{equation}
The $\textrm{GL}(2)$ integration, $\int_{\textrm{GL}(2)}\equiv\int \frac{\langle g_1 dg_1\rangle \wedge \langle g_2 dg_2 \rangle}{\langle g_1 g_2 \rangle^2}$, is as described below.

\subsection{Forward limit term}

Here, we outline how to compute the last term of (\ref{looprecursion}), known as the forward limit term. We use (\ref{lambdaABmeasure}) to rewrite the integral in terms of $\lambda_A$ and $\lambda_B$,
\begin{align}
\frac{1}{(2 \pi i)^2} \int d^{4|8}X \wedge \frac{\langle A dA\rangle\wedge \langle B dB \rangle}{\langle A B \rangle^2}\hspace{1mm} [n-1,n,1,A,B] \langle \mathcal{L}[1,\ldots,n-1,\hat{n}_{AB},\hat{\mathcal{Z}},B] \mathcal{L}[\tilde{1},\ldots,\tilde{n}_2]\rangle^{\rm c} \, .
\end{align}
where we omit the $\frac{g^2}{4\pi^2}$ factor. The $\lambda_A$ and $\lambda_B$ integration sends $\mathcal{Z}_A,\mathcal{Z}_B\rightarrow \hat{\mathcal{Z}}$. We can change variables to express the integration in terms of a GL(2) transformation of the points $\{Z_A,Z_B\}$ along the line $X=(AB)$,
\begin{align}
\frac{1}{(2 \pi i)^2} \int d^{4|8}X \wedge \frac{\langle g_1 dg_1\rangle \wedge \langle g_2 dg_2 \rangle}{\langle g_1 g_2 \rangle^2}\hspace{1mm} [n-1,n,1,A',B'] \langle \mathcal{L}[1,\ldots,n-1,\hat{n}_{AB},\hat{\mathcal{Z}},B'] \mathcal{L}[\tilde{1},\ldots,\tilde{n}_2]\rangle^{\rm c} \, .
\end{align}
where $g=(g_1 | g_2)$, $g_1=(g_{11} \hspace{1mm} g_{12})^T$, $g_2=(g_{21} \hspace{1mm} g_{22})^T$, $A'=g_{11}A+g_{12}B$, $B'= g_{21}A+g_{22}B$. The integrand has simple poles in $g_{ij}$ variables and we take the integration contour to go around the poles in $g_{ij}$, for which $\langle n_1-1\hspace{0.5mm} n_1 \hspace{0.5mm}1 \hspace{0.5mm}A'\rangle,\langle n_1-1\hspace{0.5mm} n_1 \hspace{0.5mm}1 \hspace{0.5mm}B'\rangle\rightarrow 0$. This corresponds to sending both $A'$ and $B'$ to to the point $\hat{\mathcal{Z}}=(A B)\cap(n_1-1\hspace{0.5mm} n_1\hspace{0.5mm} 1)$. Let $g_{11}^0$ and $g_{21}^0$ be the values of $g_{11}$ and $g_{21}$ which solve $\langle n_1-1 \hspace{0.5mm}n_1 \hspace{0.5mm} 1 \hspace{0.5mm} A'\rangle=0$ and $\langle n_1-1 \hspace{0.5mm}n_1 \hspace{0.5mm} 1 \hspace{0.5mm} B'\rangle=0$, respectively. Expanding the GL(2) measure and taking the residue with respect to the poles in $g_{11}$ and $g_{21}$ leaves us with the standard loop integration and fermionic integration.
\begin{align}
    & \frac{1}{(2 \pi i)^2} \int D^{3|4}\mathcal{Z}_A \wedge D^{3|4}\mathcal{Z}_B \hspace{1mm}[n-1,n,1,A,B] \langle \mathcal{L}[1,\ldots,n-1,\hat{n}_{AB},\hat{\mathcal{Z}},B] \mathcal{L}[\tilde{1},\ldots,\tilde{n}_2]\rangle^{\rm c}\notag\\
    &=- \int d^{4|8}X \lim_{g_{21}\rightarrow g_{21}^0,g_{11}\rightarrow g_{11}^0}\bigg[(g_{21}-g_{21}^0)(g_{11}-g_{11}^0)\frac{g_{12}g_{22}}{\langle g_1 g_2 \rangle^2}\notag\\
    &\hspace{40mm}\times\left([n-1,n,1,A',B'] \langle \mathcal{L}[1,\ldots,n-1,\hat{n}_{AB},\hat{\mathcal{Z}},B'] \mathcal{L}[\tilde{1},\ldots,\tilde{n}_2]\rangle^{\rm c}\right)\bigg]
\end{align}
The residue for the $g_{11}$ pole is taken first, then $g_{21}$. Dependence on $g_{12}$ and $g_{22}$ drops out upon taking the residue.

By using tree-level BCFW on the forward limit integrand in the single Wilson loop BCFW expression at $O(g^2)$, the GL(2) integral can be performed explicitly \cite{Bourjaily:2013mma}, giving a sum over terms with a Kermit piece and a product of two tree-level Wilson loops. Similarly, at higher loops, a sequence of BCFW expansions can be performed for the forward limit integrand and the GL(2) integration can be computed explicitly giving a sum over terms with a ``bridge factor" and a product of two Wilson loops of a lower loop order \cite{Bourjaily:2023apy}. We expect a similar procedure to allow for the computation of the GL(2) integral in the forward limit term for Wilson loop correlators of any loop order.

\section{Consistency checks at tree-level}
Let us first consider the tree-level recursion relation (\ref{BCFWlargeNLL}) or equivalently (\ref{looprecursion}) with $g$ set to zero. Here we first present some simple analytic checks at low MHV degree and then comment on the more extensive numerical checks which we have performed. Since the leading order contribution to the connected part of the correlation function of two Wilson loops always comes with a colour factor of \(\frac{1}{N^2}\) in the large \(N\) limit, here we omit these \(\frac{1}{N^2}\) factors.

\subsection{MHV}
In the MHV case, the connected part of the correlator is equal to zero. To see that the recursion (\ref{BCFWlargeNLL}) predicts this, note that since all of the terms on the right-hand side except for the first already come dressed with R-invariants (of Grassmann degree \(4\)), only the first term contributes, and so we see that
\begin{equation}
\langle \mathcal{L}[1,\ldots,n_1]\mathcal{L}[\tilde{1},\ldots,\tilde{n}_2]\rangle^{\rm c, tree, MHV} = \langle \mathcal{L}[1,\ldots,n_1-1]\mathcal{L}[\tilde{1},\ldots,\tilde{n}_2]\rangle^{\rm c, tree, MHV}
\end{equation}
Note that in this case there is no distinction between U($N$) and SU($N$) since the \(\alpha\)-dependent term in the recursion is already at too high an MHV degree to contribute. Iterating, we see
\begin{equation}
\langle \mathcal{L}[1,\ldots,n_1]\mathcal{L}[\tilde{1},\ldots,\tilde{n}_2]\rangle^{\rm c, tree, MHV} = \langle \mathcal{L}[1,2]\mathcal{L}[\tilde{1},\ldots,\tilde{n}_2]\rangle^{\rm c, tree, MHV}
\end{equation}
which vanishes since the first loop operator has degenerated to a backtracking Wilson loop. Thus, the recursion relation correctly predicts a vanishing contribution at MHV.

\subsection{NMHV}
Let us next consider the case of the NMHV contribution, for which we know from twistor Wilson loop diagrams that in the U($N$) theory, 
\begin{equation}
\langle \mathcal{L}[1,\ldots,n_1] \mathcal{L}[\tilde{1},\ldots,\tilde{n}_2] \rangle^{\rm c, tree, NMHV}_{{\rm U}(N)} = \sum_{i,\tilde{\jmath}}[*,i-1,i,\tilde{\jmath}-1,\tilde{\jmath}]
\label{NMHVUN}
\end{equation}
where in the sum, \(i\) runs over the cusps of the first loop and \(\tilde{\jmath}\) runs over the cusps of the second loop. In the SU($N$) theory, the answer is simply zero. 

In the SU($N$) theory, the MHV contribution to a single Wilson loop is \(1\), while for a correlator of two Wilson loops the contribution is zero. Since all terms multiplying R-invariants in (\ref{BCFWlargeNLL}) must enter at MHV in order to contribute at NMHV for the overall answer, in the SU($N$) theory the recursion reads
\begin{equation}
\langle \mathcal{L}[1,\ldots,n_1] \mathcal{L}[\tilde{1},\ldots,\tilde{n}_2] \rangle^{\rm c, tree,NMHV}_{{\rm SU}(N)} = \langle \mathcal{L}[1,\ldots,n_1-1] \mathcal{L}[\tilde{1},\ldots,\tilde{n}_2] \rangle^{\rm c, tree,NMHV}_{{\rm SU} (N)};
\end{equation}
note that the term in the final line of (\ref{BCFWlargeNLL}) has vanished since for SU($N$) we have a factor of \(1-\alpha = 0\) multiplying the R-invariant. Iterating, this correctly predicts a vanishing contribution at NMHV by the same logic as we used in the case of the MHV contribution. 

In U($N$), the final term in the recursion survives and we instead find
\begin{align}
\langle \mathcal{L}[1,\ldots,n_1] \mathcal{L}[\tilde{1},\ldots,\tilde{n}_2] \rangle^{\rm c,tree, NMHV}_{{\rm U}(N)} = &\langle \mathcal{L}[1,\ldots,n_1-1] \mathcal{L}[\tilde{1},\ldots,\tilde{n}_2] \rangle^{\rm c, tree, NMHV}_{{\rm U}(N)}\notag \\
&+\sum_{\tilde{\jmath}}[n_1-1,n_1,1,\tilde{\jmath}-1,\tilde{\jmath}]
\end{align}
where again we have used that any object multiplying an R-invariant must enter at MHV in order to give an NMHV contribution overall. By using (\ref{NMHVUN}) and identifying the reference twistor with \(Z_1\), the right-hand side reads
\begin{equation}
\sum_{i=1}^{n_1-1} \sum_{\tilde{\jmath}=\tilde{1}}^{\tilde{n}_2} [1, i-1, i, \tilde{\jmath}-1,\tilde{\jmath}] + \sum_{\tilde{\jmath}}[n_1-1,n_1,1,\tilde{\jmath}-1,\tilde{\jmath}].
\end{equation}
The second term is precisely the \(i=n_1\) contribution to the first sum and so, combining, we have
\begin{equation}
\sum_{i=1}^{n_1} \sum_{\tilde{\jmath}=\tilde{1}}^{\tilde{n}_2} [1, i-1, i, \tilde{\jmath}-1,\tilde{\jmath}]
\end{equation}
which is the correct formula. 

\subsection{Discussion of the `intersecting' term}
The N\(^2\)MHV contribution is the first time that the various Wilson loop correlators which feature in the recursion (other than the first, boundary term) actually enter non-trivially beyond their MHV expressions. Let us briefly comment on the novel term
\begin{equation}
[n-1,n,1,\tilde{\jmath}-1,\tilde{\jmath}]\langle \mathcal{L}[1,\ldots,\hat{n}_1,I_{\tilde{\jmath}},\tilde{\jmath},\ldots,\tilde{\jmath}-1,I_{\tilde{\jmath}}]\rangle
\end{equation}
which is really a Wilson loop in a self-intersecting configuration: the arguments first run around the first Wilson loop to the intersection point \(I_{\tilde{\jmath}}\), then around the edges of the second Wilson loop and back to \(I_{\tilde{\jmath}}\). A Wilson loop in such a self-intersecting limit transpires to be divergent, but the presence of the R-invariant \([n-1,n,1,\tilde{\jmath}-1,\tilde{\jmath}]\) eliminates the divergences for Grassmann reasons and protects the well-definedness of this term in the recursion relation. 

As a concrete example, let us consider the intersecting Wilson loop 
\begin{equation}
\langle \mathcal{L}[1,2,3,\widehat{4}_1,I_{\tilde{1}},\tilde{2},\tilde{3},\tilde{4},I_{\tilde{1}}] \rangle
\end{equation}
and for simplicity specialise to the NMHV contribution to this object (which in turn contributes to the connected part of the correlator of two Wilson loops at N\(^2\)MHV). If we simply take the twistor-diagrammatic formula 
\begin{equation}
\sum_{i<j} [*,i-1,i,j-1,j]
\end{equation}
and substitute in naively for the twistors in our intersecting Wilson loop, two issues arise: 
\begin{enumerate}
\item Due to the repetition of the twistor \(I_{\tilde{1}}\) in the loop, we will have R-invariants with a repeated twistor, e.g. \([*,I_{\tilde{1}},\tilde{1},\tilde{4},I_{\tilde{1}}]\). Such terms are ill-defined since the antisymmetry of R-invariants dictates that they vanish while they clearly also have a zero factor in their denominator.

These terms can be addressed in a simple way: if we first perturb the second instance of the twistor \(I_{\tilde{\jmath}}\), i.e. instead consider Wilson loop
\begin{equation}
\langle \mathcal{L}[1,2,3,\widehat{4}_1,I_{\tilde{1}},\tilde{2},\tilde{3},\tilde{4},I'_{\tilde{1}}] \rangle
\end{equation}
with
\begin{equation}
\mathcal{Z}_{I'_{\tilde{1}}} = \mathcal{Z}_{I_{\tilde{1}}} + \epsilon \mathcal{Z}_r
\end{equation}
and then take the limit \(\epsilon \to 0\) in a way which respects \(Q\)-supersymmetry (i.e. which treats the \(Z\) and \(\chi\) variables in the same way), by power-counting these terms are \(\epsilon\)-suppressed and so are identified with zero in the limit. 

\item The specific geometry of the Wilson loop in this configuration - namely, the fact that \(4\), \(I_{\tilde{1}}\) and \(1\) are collinear, as are \(\tilde{1}\), \(I_{\tilde{1}}\) and \(\tilde{4}\) - gives rise to other terms which are divergent even though they don't feature repeated twistors. Namely, we have for this case 
\begin{equation}
A = [*,1,2,\tilde{4}_1,I_{\tilde{1}}], \, \, \, \, B = [*,I_{\tilde{1}},3,\hat{4}_1], \, \, \, \ C = [*,I_{\tilde{1}},\tilde{1},\tilde{3},\tilde{4}], \, \, \, \, D = [*,\tilde{4},I_{\tilde{1}},\tilde{1},\tilde{2}].
\end{equation}

For instance, \(A\) features the denominator factor
\begin{equation}
\langle * \, 1 \,\tilde{4}_1 \,I_{\tilde{1}} \rangle = \langle *\, 1\, (34)\cap(\tilde{4}\tilde{1}1) \,(\tilde{4} \tilde{1}) \cap (341) \rangle 
\end{equation}
which vanishes on support of Plücker relations. If one regulates these diagrams in the same way as the previous case, by replacing \(Z_{I_{\tilde{1}}}\) with 
\begin{equation}
Z_{I'_{\tilde{1}}} = Z_{I_{\tilde{1}}} + \epsilon Z_r
\end{equation}
and then taking the \(\epsilon \to 0\) limit supersymmetrically, \(A\) and \(B\) are \(\epsilon\)-suppressed and so vanish in the limit, while \(C\) and \(D\) in fact diverge.

However, the Grassmann structure of \(C\) and \(D\) degenerates in the \(\epsilon \to 0\) limit and as such becomes a linear combination of \(\chi_3\), \(\chi_4\), \(\chi_1\), \(\chi_{\tilde{4}}\) and \(\chi_{\tilde{1}}\). Note that these are precisely the arguments of the R-invariant \([3,4,1,\tilde{4},\tilde{1}]\) which accompany the R-invariant in the recursion. As such, those terms which survive in the limit and naively diverge as \(\epsilon \to 0\) are actually eliminated for Grassmann reasons. The limit-taking prescription thus allows us to obtain a perfectly well-defined object, \emph{provided that we multiply by the R-invariant before taking the limit}.

Note that this is exactly the same phenomenon as was already present in the 'forward limit' term in the all-loop recursion relation for a single Wilson loop presented in \cite{Arkani-Hamed:2010zjl}.

\end{enumerate}

As such, the self-intersecting term should really be understood in the following way: we consider
\begin{equation}
[n-1,n,1,\tilde{\jmath}-1,\tilde{\jmath}]\langle \mathcal{L}[1,\ldots,\hat{n}_1,I'_{\tilde{\jmath}},\tilde{\jmath},\ldots,\tilde{\jmath}-1,I''_{\tilde{\jmath}}]\rangle
\end{equation}
where
\begin{equation}
\mathcal{Z}_{I'_{\tilde{\jmath}}} =  \mathcal{Z}_{I_{\tilde{\jmath}}}  + \epsilon \mathcal{Z}_r, \, \, \, \, \, \, \, \, \mathcal{Z}_{I''_{\tilde{\jmath}}} =  \mathcal{Z}_{I_{\tilde{\jmath}}}  + \eta \mathcal{Z}_s
\end{equation}
for arbitrary super-twistors \(\mathcal{Z}_r\) and \(\mathcal{Z}_s\), and then take the limit \(\epsilon \to 0\) followed by \(\eta \to 0\) (or equivalently \(\eta \to 0\) followed by \(\epsilon \to 0\)). Any apparent divergences come with vanishing Grassmann factors and so may be discarded, and we are left with a finite, well-defined object. Note that although we have considered the NMHV contribution for the purposes of a simple example, this prescription continues to work at higher MHV degree. 

Operationally, if working with twistor Wilson loop diagrams, it is sufficient to first generate an ordinary Wilson loop with the appropriate number of edges, then plug in the values of the various twistors (including e.g. the repetition of \(I_{\tilde{\jmath}}\)), before discarding any diagrams which are naively divergent. Doing this will give an ill-defined answer in general (for instance, it is easy to check that the answer isn't even independent of the reference supertwistor \(\mathcal{Z}_*\)), since we are discarding terms which really diverge in the appropriate limit, but upon multiplying by the accompanying R-invariant we obtain the correct, reference twistor independent contribution. 

\subsection{N\texorpdfstring{\(^2\)}{2}MHV in the SU(\texorpdfstring{$N$}{N}) theory}

Specialising to the SU($N$) theory, let us now turn to the N\(^2\)MHV contribution to the connected part of the correlator of two Wilson loops. As has already been noted, this object remarkably factorises into a perfect square: 
\begin{equation}
\langle\mathcal{L}[1,\ldots,n_1]\mathcal{L}[\tilde{1},\ldots,\tilde{n}_1] \rangle^{\rm{c,tree, N}^2\rm{MHV}}_{{\rm SU}(N)}= \frac{1}{2}\Bigl(\langle\mathcal{L}[1,\ldots,n_1]\mathcal{L}[\tilde{1},\ldots,\tilde{n}_2] \rangle^{\rm{c, tree, NMHV}}_{{\rm U}(N)} \Bigr)^2.
\end{equation}

However, it is not at all manifest from the BCFW representation that the answer factorises in this way, and we comment that it requires a surprisingly involved and tedious calculation to see this. Here we present an analytic proof of the equivalence of the two expressions.

\subsubsection{\texorpdfstring{\(n_1=3\)}{n1=3}}

Let us begin with the simple case of \(n_1 = 3\) with generic \(n_2\). In this case (\ref{BCFWlargeNLL}) reads (here all terms are at tree-level and computed in the SU($N$) theory)
\begin{align}
\langle \mathcal{L}&(1,2,3)\mathcal{L}[\tilde{1},\ldots,\tilde{n}_2]\rangle^{\rm{c, N}^2\rm{MHV}} \notag \\
&= \sum_{\tilde{\jmath}}[1,2,3,\tilde{\jmath}-1,\tilde{\jmath}]\Bigl(\langle \mathcal{L}[1,2,\hat{3},I_{\tilde{\jmath}},\tilde{\jmath},\ldots,\tilde{\jmath}-1,I_{\tilde{\jmath}}]\rangle^{\rm NMHV} - \langle \mathcal{L}[\tilde{1},\ldots,\tilde{n}_2]\rangle\Bigr)
\end{align}
where we recall that
\begin{equation}
\widehat{\mathcal{Z}}_3 = \mathcal{Z}_2 \langle 3 \, 1 \, \tilde{\jmath}-1 \tilde{\jmath} \rangle- \mathcal{Z}_3 \langle 2 \, 1 \, \tilde{\jmath}-1 \tilde{\jmath} \rangle , \, \, \, \, \, \, \mathcal{Z}_{I_{\tilde{\jmath}}} = \mathcal{Z}_{\tilde{\jmath}-1} \langle \tilde{\jmath} \, 1 \, 2 \, 3 \rangle - \mathcal{Z}_{\tilde{\jmath}} \langle \tilde{\jmath} -1 \, 1 \, 2 \, 3 \rangle.
\end{equation}

Choosing \(\mathcal{Z}_* = \mathcal{Z}_2\), and recalling that in our prescription any R-invariants with a repeated twistor are identified with zero, the expression in the brackets on the right-hand side reads (here we use the twistor Wilson loop diagram expansion for the objects on the right) 
\begin{align}
&\sum_{\tilde{k}=\tilde{\jmath}+1}^{\tilde{\jmath}-1}[2,\hat{3},I_{\tilde{\jmath}},\tilde{k}-1, \tilde{k}] + \sum_{\tilde{k}=\tilde{\jmath}+2}^{\tilde{\jmath}-1}[2,I_{\tilde{\jmath}},\tilde{\jmath},\tilde{k}-1, \tilde{k}]
+ \sum_{\tilde{k}=\tilde{\jmath}+1}^{\tilde{\jmath}-2}[2,\tilde{k}-1,\tilde{k},\tilde{\jmath}-1,\tilde{\jmath}] \notag \\
& +\sum_{\tilde{k}=\tilde{\jmath}+1}^{\tilde{\jmath}-1}[2,\tilde{k}-1,\tilde{k},I_{\tilde{\jmath}},1] - \sum_{\tilde{\jmath}+2}^{\tilde{\jmath}-2}[2,\tilde{\jmath}-1,\tilde{\jmath},\tilde{k}-1,\tilde{k}]
\label{theSum}
\end{align}
after some cancellation. To simplify the analysis, we can use the invariance of the object under \(Q^A_{A'} = \sum Z_i^A \frac{\partial}{\partial \chi_i^{A'}}\), to set up to four \(\chi\) variables to zero. Let us choose to set \(\chi_1 = \chi_2 = \chi_3 = 0\), from which it follows that we also have \(\chi_{\hat{3}} = 0\). Simplifying the R-invariant factor in the front of the right-hand side, we have after some calculation that
\begin{equation}
[1,2,3,\tilde{\jmath}-1,\tilde{\jmath}] = \frac{\bar{\delta}^{0|4}(\chi_{I_{\tilde{\jmath}}})}{\langle 1\, 2 \, 3 \tilde{\jmath}-1\rangle \langle 2 \, 3 \, \tilde{\jmath}-1 \, \tilde{\jmath} \rangle \langle 3 \, \tilde{\jmath}-1 \, \tilde{\jmath} \, 1\rangle \langle \tilde{\jmath}-1 \, \tilde{\jmath} \, 1 \, 2 \rangle \langle \tilde{\jmath} \, 1 \, 2 \, 3 \rangle }
\end{equation}

Simplifying the other R-invariants on the support of the \(\bar{\delta}^{0|4}(\chi_{I_{\tilde{\jmath}}})\) which by the above is multiplying them all (and recalling that we have set \(\chi_1,\chi_2,\chi_3\) to zero), it follows from some simple algebra that
\small{
\begin{align}
&[2,\hat{3},I_{\tilde{\jmath}},\tilde{k}-1,\tilde{k}] \notag \\
&= -\frac{\langle 1 \, 2 \, \tilde{\jmath}-1 \, \tilde{\jmath}\rangle \langle \tilde{\jmath}-1 \, \tilde{\jmath} \, 2 \, 3 \rangle \bar{\delta}^{0|4}\bigl(\chi_{\tilde{k}-1}\langle \tilde{k} \, 1 \, 2 \, 3 \rangle - \chi_{\tilde{k}}\langle \tilde{k}-1 \, 1 \, 2 \, 3\rangle\bigr)}{\langle \tilde{k}-1 \, \tilde{k} \, 2 \, 3\rangle \langle \tilde{k} \, 1 \, 2 \, 3 \rangle \langle \tilde{k} -1 \, 1 \, 2 \, 3 \rangle \langle 123[\tilde{\jmath}-1\rangle \langle \tilde{\jmath}] \, 1\, \tilde{k}-1 \, \tilde{k} \rangle  \langle 1 \, 2 \, 3[\tilde{\jmath}-1\rangle \langle \tilde{\jmath}] \, \tilde{k}-1 \, \tilde{k} \, 2 \rangle},
\end{align}}

\begin{align}
&[2, I_{\tilde{\jmath}}, \tilde{\jmath}, \tilde{k}-1, \tilde{k}] \notag \\
&= \frac{\langle \tilde{\jmath} \, 1 \, 2 \, 3 \rangle \bar{\delta}^{0|4}\bigl(\chi_{\tilde{\jmath}-1}\langle \tilde{\jmath}  \, \tilde{k}-1 \, \tilde{k} \, 2 \rangle - (\tilde{\jmath}-1 \leftrightarrow \tilde{\jmath}) + (\tilde{\jmath} \leftrightarrow \tilde{k})\bigr)}{\langle 2 \, \tilde{\jmath} -1 \, \tilde{\jmath} \, \tilde{k} -1 \rangle \langle \tilde{\jmath}-1 \, \tilde{\jmath} \, \tilde{k}-1 \, \tilde{k} \rangle \langle \tilde{\jmath} \, \tilde{k}-1 \, \tilde{k} \, 2 \rangle \langle \tilde{k}-1 \, \tilde{k} \, 2 [\tilde{\jmath} -1 \rangle \langle \tilde{\jmath}] \, 1 \, 2 \, 3 \rangle \langle \tilde{k} \, 2 \, \tilde{\jmath}-1 \, \tilde{\jmath} \rangle },
\end{align}

\begin{align}
&[2, \tilde{k}-1, \tilde{k},\tilde{\jmath},I_{\tilde{\jmath}}] \notag \\
&=-\frac{\langle \tilde{\jmath}-1 \, 1 \, 2 \, 3 \rangle\bar{\delta}^{0|4}\bigl(\chi_{\tilde{\jmath}-1} \langle \tilde{\jmath}  \, \tilde{k}-1 \, \tilde{k} \, 2\rangle - (\tilde{\jmath}-1 \leftrightarrow \tilde{\jmath}) + (\tilde{\jmath} \leftrightarrow \tilde{k})\bigr)}{ \langle 2 \, \tilde{k}-1 \, \tilde{k} \, \tilde{\jmath} -1 \rangle \langle \tilde{k}-1 \, \tilde{k} \, \tilde{\jmath}-1 \, \tilde{\jmath} \rangle \langle \tilde{k} \, \tilde{\jmath}-1 \, \tilde{\jmath} \, 2 \rangle \langle \tilde{\jmath}-1 \, \tilde{\jmath} \, 2 \, \tilde{k}-1 \rangle \langle 1 \, 2 \, 3 \, [\tilde{\jmath}-1 \rangle \langle \tilde{\jmath}] \, 2 \, \tilde{k}-1 \, \tilde{k} \rangle  },
\end{align}

\begin{align}
&[2, \tilde{k}-1, \tilde{k}, I_{\tilde{\jmath}}, 1] \notag \\
&=\frac{\langle 1 \, 2 \, \tilde{\jmath}-1 \, \tilde{\jmath} \rangle^2 \bar{\delta}^{0|4}\bigl(\chi_{\tilde{k}-1} \langle \tilde{k} \, 1 \, 2 \, 3 \rangle - \chi_{\tilde{k}} \langle \tilde{k}-1 \, 1 \, 2 \, 3 \rangle \bigr)}{\langle 2 \, \tilde{k}-1 \, \tilde{k} \, 1 \, [ \tilde{\jmath}-1 \rangle \langle \tilde{\jmath}]\, 1 \, 2 \, 3 \rangle \langle \tilde{k}-1 \, \tilde{k} \, 1 \, [ \tilde{j-1} \rangle \langle \tilde{\jmath}] \, 1 \, 2 \, 3 \rangle \langle \tilde{k} \, 1 \, 2 \, 3 \rangle \langle \tilde{k}-1 \, 1 \, 2 \, 3 \rangle \langle 1 \, 2 \, \tilde{k}-1 \, \tilde{k} \rangle},
\end{align}

and finally

\begin{align}
&[2, \tilde{\jmath}-1, \tilde{\jmath}, \tilde{k}-1, \tilde{k}] \notag \\
&= -\frac{\bar{\delta}^{0|4}\bigl(\chi_{\tilde{\jmath}-1} \langle \tilde{\jmath}  \, \tilde{k}-1 \, \tilde{k} \, 2\rangle - (\tilde{\jmath}-1 \leftrightarrow \tilde{\jmath}) + (\tilde{\jmath} \leftrightarrow \tilde{k})\bigr)}{\langle 2 \, \tilde{\jmath}-1 \, \tilde{\jmath} \, \tilde{k}-1 \rangle \langle \tilde{\jmath}-1 \, \tilde{\jmath} \, \tilde{k}-1 \, \tilde{k} \rangle \langle \tilde{\jmath} \, \tilde{k}-1 \, \tilde{k} \, 2 \rangle \langle \tilde{k}-1 \, \tilde{k} \, 2 \, \tilde{\jmath}-1 \rangle \langle \tilde{k} \,2 \, \tilde{\jmath}-1 \, \tilde{\jmath} \rangle}.
\end{align}

Note in particular that the Grassmann structure has collapsed into just two terms,
\begin{equation}
G_1 = \bar{\delta}^{0|4}\bigl(\chi_{\tilde{\jmath}-1} \langle \tilde{\jmath}  \, \tilde{k}-1 \, \tilde{k} \, 2\rangle - (\tilde{\jmath}-1 \leftrightarrow \tilde{\jmath}) + (\tilde{\jmath} \leftrightarrow \tilde{k})\bigr)
\end{equation}
and
\begin{equation}
G_2 = \bar{\delta}^{0|4}\bigl(\chi_{\tilde{k}-1} \langle \tilde{k} \, 1 \, 2 \, 3 \rangle - \chi_{\tilde{k}} \langle \tilde{k}-1 \, 1 \, 2 \, 3 \rangle \bigr)
\end{equation}

Performing the summation in (\ref{theSum}), and noting that all of the sums may be taken to run from \(\tilde{\jmath}+2\) to \(\tilde{\jmath}-2\) at no cost since
\begin{equation}
\chi_{\tilde{\jmath}-1}\langle \tilde{\jmath} \, \tilde{k}-1 \, \tilde{k} \, 2 \rangle - \chi_{\tilde{\jmath}} \langle \tilde{\jmath} - 1 \, \tilde{k}-1 \, \tilde{k} \, 2 \rangle + \chi_{\tilde{k}-1}\langle \tilde{k} \, \tilde{\jmath} -1 \, \tilde{\jmath}  \, 2 \rangle - \chi_{\tilde{k}} \langle \tilde{k}-1 \, \tilde{\jmath}-1 \, \tilde{\jmath} \, 2 \rangle = 0 
\end{equation}
for \(\tilde{k}=j+1\) or \(\tilde{k}=j-1\), we find that the coefficient of \(G_1\) vanishes while the coefficient of \(G_2\) reduces to 
\begin{equation}
\frac{\langle 1 \, 2 \, \tilde{\jmath}-1 \, \tilde{\jmath} \rangle}{\langle \tilde{k} \, 1 \, 2 \, 3 \rangle \langle \tilde{k} -1 \, 1 \, 2 \, 3 \rangle \langle 1 \, 2 \, 3 \, [\tilde{\jmath}-1 \rangle \langle \tilde{\jmath}] \, 1 \, \tilde{k}-1 \, \tilde{k} \rangle \langle 1 \, 2 \, 3 [ \tilde{\jmath}-1 \rangle \langle j] \, 2 \, \tilde{k}-1 \, \tilde{k} \rangle} \times \Bigl( -\frac{\langle \tilde{\jmath}-1 \, \tilde{\jmath} \, 2 \, 3 \rangle}{\langle \tilde{k}-1 \, \tilde{k} \, 2 \, 3 \rangle}  + \frac{\langle \tilde{\jmath}-1 \, \tilde{\jmath} \, 1 \, 2 \rangle}{\langle \tilde{k}-1 \, \tilde{k} \, 1 \, 2 \rangle} \Bigr)
\end{equation}
Recalling that we have set \(\chi_1=\chi_2=\chi_3 = 0\), \(G_2\) is precisely the numerator of \([1,2,3,\tilde{k}-1,\tilde{k}]\), while its coefficient simplifies to \(\frac{1}{2}\) times the denominator of that R-invariant. Thus, the overall expression reduces to 
\begin{equation}
\frac{1}{2}[1,2,3,\tilde{k}-1,\tilde{k}] 
\end{equation}

Finally, restoring the R-invariant prefactor and the sum over \(\tilde{\jmath}\), we are left with
\begin{equation}
\frac{1}{2}\sum_{\tilde{\jmath}}[1,2,3,\tilde{\jmath}-1,\tilde{\jmath}]\sum_{\tilde{k} \neq \tilde{\jmath}}[1,2,3,\tilde{k}-1,\tilde{k}]
\end{equation}
\begin{equation}
= \frac{1}{2}\bigl(\sum_{\tilde{\jmath}}[1,2,3,\tilde{\jmath}-1,\tilde{\jmath}]\bigr)^2
\end{equation}
where we have used that R-invariants are nilpotent in the final line. This exactly matches \(\frac{1}{2}\) the square of the NMHV tree contribution to the correlator of a triangle with an \(n\)-gon in the U($N$) theory.

\subsubsection{General \texorpdfstring{\(n_1\)}{n1}}
Let us now extend the analysis to general \(n_1\); the recursion relation states that 
\begin{align}
\langle \mathcal{L}[1,\ldots,n_1]&\mathcal{L}[\tilde{1},\ldots,\tilde{n}_2]\rangle^{\rm{c,N}^2\rm{MHV}} = \langle \mathcal{L}[1,\ldots,n_1-1]\mathcal{L}[\tilde{1},\ldots,\tilde{n}_2]\rangle^{\rm{c,N}^2\rm{MHV}} \notag \\
+\sum_{\tilde{\jmath}}&[1,n_1-1,n_1,\tilde{\jmath}-1,\tilde{\jmath}]\Bigl( \langle \mathcal{L}[1,\ldots,n_1-1,\hat{n}_{1,j}, I_{\tilde{\jmath}},\tilde{\jmath},\ldots,\tilde{\jmath}-1,I_{\tilde{\jmath}}] \rangle^{\rm NMHV}  \notag \\
-& \langle \mathcal{L}[1,\ldots,\hat{n}_1]\rangle^{\rm NMHV} - \langle \mathcal{L}[\tilde{1},\ldots,\hat{n}_2]\rangle^{\rm NMHV} \Bigr).
\end{align}

Again expanding all of the objects on the right-hand side diagrammatically, we obtain after some cancellation of terms that (here we set \(\mathcal{Z}_* = \mathcal{Z}_{n_1-1}\) which also eliminates some terms)
\begin{align}
 &\langle \mathcal{L}[1,\ldots,n_1-1,\hat{n}_{1,j}, I_{\tilde{\jmath}},\tilde{\jmath},\ldots,\tilde{\jmath}-1,I_{\tilde{\jmath}}] \rangle^{\rm NMHV} - \langle \mathcal{L}[1,\ldots,\hat{n}_1]\rangle^{\rm NMHV} - \langle \mathcal{L}[\tilde{1},\ldots,\hat{n}_2]\rangle^{\rm NMHV} \notag \\
=&
\sum_{i=2}^{n_1-2}[n_1-1, i-1,i,\hat{n}_1,I_{\tilde{\jmath}}] + \sum_{i=2}^{n_1-2}[n_1-1,i-1,i,I_{\tilde{\jmath}},\tilde{\jmath}] + \sum_{i=2}^{n_1-2}\sum_{\tilde{l}=\tilde{\jmath}+1}^{\tilde{\jmath}-1}[n_1-1,i-1,i,\tilde{l}-1,\tilde{l}] \notag \\
&+ \sum_{i=2}^{n_1-2}[n_1-1,i-1,i,\tilde{\jmath}-1,I_{\tilde{\jmath}}] +\sum_{i=2}^{n_1-2}[n_1-1,i-1,i,I_{\tilde{\jmath}},1] - \sum_{i=2}^{n_1-2}[n_1-1,i-1,i,\hat{n}_1,1] \notag \\&+\sum_{\tilde{l}=\tilde{\jmath}+1}^{\tilde{\jmath}-1}[n_1-1,\hat{n}_1,I_{\tilde{\jmath}},\tilde{l}-1,\tilde{l}] +\sum_{\tilde{l}=\tilde{\jmath}+1}^{\tilde{\jmath}-1}[n_1-1,I_{\tilde{\jmath}},\tilde{\jmath},\tilde{l}-1,\tilde{l}] + \sum_{\tilde{l}=\tilde{\jmath}+1}^{\tilde{\jmath}-2}[n_1-1,\tilde{l}-1,\tilde{l},\tilde{\jmath}-1,I_{\tilde{\jmath}}]\notag \\
& + \sum_{\tilde{l}=\tilde{\jmath}+1}^{\tilde{\jmath}-1}[n_1-1,\tilde{l}-1,\tilde{l},I_{\tilde{\jmath}},1] - \sum_{\tilde{l} = \tilde{\jmath}+2}^{\tilde{\jmath}-1}[n_1-1,\tilde{\jmath}-1,\tilde{\jmath},\tilde{l}-1,\tilde{l}].
\end{align}

After multiplying by \([1,n_1-1,n_1,\tilde{\jmath}-1,\tilde{\jmath}]\) and summing over \(\tilde{\jmath}\), the final five terms here collapse to
\begin{equation}
\frac{1}{2}\Bigl(\sum_{\tilde{\jmath}}[1,n_1-1,n_1,\tilde{\jmath}-1,\tilde{\jmath}]  \Bigr)^2
\end{equation}
via the same calculation as was presented in the \(n_1=3\) case. In order to supply all of the additional diagrams which arise for the case of \(n_1\) versus \(n_1-1\), we additionally need
\begin{equation}
\sum_{\tilde{\jmath}}[1,n_1-1,n_1,\tilde{\jmath}-1,\tilde{\jmath}] \sum_{i=2}^{n_1-2}\sum_{\tilde{l}}[n_1-1,i-1,i,\tilde{l}-1,\tilde{l}]
\end{equation}
and so completing the proof amounts to establishing that
\begin{align}
&
\sum_{i=2}^{n_1-2}[n_1-1, i-1,i,\hat{n}_1,I_{\tilde{\jmath}}] + \sum_{i=2}^{n_1-2}[n_1-1,i-1,i,I_{\tilde{\jmath}},\tilde{\jmath}] + \sum_{i=2}^{n_1-2}\sum_{\tilde{l}=\tilde{\jmath}+1}^{\tilde{\jmath}-1}[n_1-1,i-1,i,\tilde{l}-1,\tilde{l}] \notag \\
&+ \sum_{i=2}^{n_1-2}[n_1-1,i-1,i,\tilde{\jmath}-1,I_{\tilde{\jmath}}] +\sum_{i=2}^{n_1-2}[n_1-1,i-1,i,I_{\tilde{\jmath}},1] - \sum_{i=2}^{n_1-2}[n_1-1,i-1,i,\hat{n}_1,1] \notag \\
&=  \sum_{i=2}^{n_1-2}\sum_{\tilde{l}}[n_1-1,i-1,i,\tilde{l}-1,\tilde{l}]
\end{align}

The third term on the left-hand side is almost exactly equal to the right-hand side, except for the fact that it omits the \(\tilde{l}=\tilde{\jmath}\) term. We are therefore done provided that
\begin{align}
\sum_{i=2}^{n_1-2}&\Bigl([n_1-1, i-1,i,\hat{n}_1,I_{\tilde{\jmath}}] + [n_1-1,i-1,i,I_{\tilde{\jmath}},\tilde{\jmath}] -[n_1-1,i-1,i,\tilde{\jmath}-1,\tilde{\jmath}] \notag \\
&+ [n_1-1,i-1,i,\tilde{\jmath}-1,I_{\tilde{\jmath}}] +[n_1-1,i-1,i,I_{\tilde{\jmath}},1] - [n_1-1,i-1,i,\hat{n}_1,1]\Bigr) = 0
\end{align}
A simple calculation confirms that, for each \(i\), the first three and final three terms each sum to zero individually, and thus we have established the required result. 

We comment that this was a surprisingly involved check, although it is worth remembering that even the twistor diagrammatic calculation provides an expression for the N\(^2\)MHV tree correlator in terms of shifted twistors, and even in that case it is only after the use of R-invariant identities that the perfect square form arises. 

\subsection{Higher MHV degree checks at tree-level}

In addition to the analytic checks presented here, we have verified this relation numerically at NMHV, N$^2$MHV and N$^3$MHV for all correlators of two Wilson loops up to octagon-octagon, and also at N$^4$MHV for $n$-gon-square correlators up to $n=8$, for both U$(N)$ and SU$(N)$. To do this, we computed the left-hand side and right-hand side of the recursion using twistor Wilson loop diagrams and plugged in random numerics on a variety of chi monomials to verify a match.

\section{Consistency checks at one and two loops}

\subsection{MHV one-loop}

\subsubsection{SU(\texorpdfstring{$N$}{N})}
In the SU($N$) theory, the MHV contribution at O($g^2$) to the correlator of two Wilson loops vanishes, since all of the possible twistor Wilson loop diagrams come with a vanishing colour factor. This clearly matches the prediction from (\ref{looprecursion}) since only the final term has the potential to enter at this MHV degree and it vanishes on account of the fact that the tree-level connected correlator would contribute at NMHV (note that the fermionic integration removes eight Grassmann degrees) which is a zero contribution in the SU($N$) theory. Thus, only the boundary term survives and see by iterating that every correlator is equal to the vanishing backtracking loop with \(n_1=2\).

\subsubsection{U(\texorpdfstring{$N$}{N})}
In the U($N$) theory, it was shown in \cite{Drummond:2025ulh} that at O(g$^2$), the contribution to the correlator of two Wilson loops is given by 
\begin{equation}
    \sum_{i,\bar{j}}K_{i \tilde{\jmath}}
\end{equation}
for `Kermit diagrams' $K_{ij}$ defined by
\begin{equation}
K_{ij} = -\int \frac{d^4 x_{AB}}{\pi^2} \frac{(\langle *\, i-1\, i \,[A\rangle \langle B] \,j-1 \, j\, * \rangle)^2}{\langle A\,B\, i-1\, i\rangle \langle A\,B\, j-1\, j\rangle \langle A\, B\, i-1 \,* \rangle \langle A\, B\, i \,* \rangle \langle A\, B\, j-1 \,* \rangle \langle A\, B\, j \,* \rangle}\,.
\end{equation}
For concreteness, let us choose to set the arbitrary reference twistor as \(\mathcal{Z}_* = \mathcal{Z}_1\) for which the sum reduces to 
\begin{equation}
\sum_{i=3}^{n_1}\sum_{\tilde{\jmath}=\tilde{1}}^{\tilde{n}_2} K_{i\tilde{\jmath}}
\end{equation}

On the other hand, at this loop order and MHV degree the recursion (\ref{looprecursion}) reads
\begin{align}
    &\langle \mathcal{L}[1,\ldots,n_1] \mathcal{L}[\tilde{1},\ldots,\tilde{n}_2]\rangle^{\textrm{c},O(g^2),\textrm{MHV}} = \langle \mathcal{L}[1,\ldots,n_1-1] \mathcal{L}[\tilde{1},\ldots,\tilde{n}_2]\rangle^{\textrm{c},O(g^2),\textrm{MHV}} \notag \\
   &\quad +\frac{g^2}{\pi^2} \frac{1}{(2 \pi i)^2} \int D^{3|4}\mathcal{Z}_A \wedge D^{3|4}\mathcal{Z}_B [n_1-1,n_1,1,A,B] \langle \mathcal{L}[1,\ldots,n_1-1,\hat{n}_{AB},\hat{\mathcal{Z}},B] \mathcal{L}[\tilde{1},\ldots,\tilde{n}_2]\rangle^{\textrm{c,tree,NMHV}} \,.
\end{align}
Note that the fermionic integration on the second line reduces the power of the Grassmann variables from eight to zero as required at this MHV degree. Expanding the tree-level correlator diagrammatically with the choice \(\mathcal{Z}_* = \mathcal{Z}_1\) and keeping only those terms which survive the fermionic integration, the
recursion (\ref{looprecursion}) reads
\begin{align}
    &\langle \mathcal{L}[1,\ldots,n_1] \mathcal{L}[\tilde{1},\ldots,\tilde{n}_2]\rangle^{\textrm{c},O(g^2),\textrm{MHV}}= \langle \mathcal{L}[1,\ldots,n_1-1] \mathcal{L}[\tilde{1},\ldots,\tilde{n}_2]\rangle^{\textrm{c},O(g^2),\textrm{MHV}}  \notag \\
   &\quad +\frac{g^2}{\pi^2} \int D^{3|4}\mathcal{Z}_A \wedge D^{3|4}\mathcal{Z}_B [n-1,n,1,A,B][1,\hat{\mathcal{Z}},B,\tilde{\jmath}-1,\tilde{\jmath}].
\end{align}
It was shown in \cite{Arkani-Hamed:2010zjl} that the product of two R-invariants becomes precisely \(K_{n_1 \tilde{\jmath}}\) (with the choice \(\mathcal{Z}_* = \mathcal{Z}_1\)), which strictly speaking requires absorbing the \(\frac{1}{\pi^2}\) into the integration measure, so that the BCFW expression becomes
\begin{align}
    &\langle \mathcal{L}[1,\ldots,n_1] \mathcal{L}[\tilde{1},\ldots,\tilde{n}_2]\rangle^{\textrm{c},O(g^2),\textrm{MHV}}  = \langle \mathcal{L}[1,\ldots,n_1-1] \mathcal{L}[\tilde{1},\ldots,\tilde{n}_2]\rangle^{\textrm{c},O(g^2),\textrm{MHV}}  \notag \\
   &\quad +g^2 \sum_{\tilde{\jmath}}K_{n_1 \tilde{\jmath}}
\end{align}
In the case \(n_1 = 3\) this clearly matches the diagrammatic expansion (since the boundary term on the RHS degenerates to a backtracking loop), and it is similarly clear that \(g^2 \sum_{\tilde{\jmath}}K_{n_1 \tilde{\jmath}}\) provides precisely those extra diagrams which arise when increasing \(n_1-1 \to n_1\). The BCFW representation thus matches the diagrammatic representation for all \(n_1\).

\subsection{Two-loop and other one-loop checks}

We have verified the loop-level relation numerically at $N$MHV for all $O(g^2)$ correlators of two Wilson loops up to octagon-octagon and at N$^2$MHV for $n$-gon-square up to $n=6$ for U$(N)$ and SU$(N)$. We also checked the relation for MHV and NMHV correlators at $O(g^4)$ up to the case of a pentagon-square correlator. As in the case of our tree-level checks, these were performed by using the diagrammatic expansion for the correlators appearing on both sides of the equation and evaluating each side of the recursion on various chi monomials for random numerics. 

\section{Recursion beyond the large \texorpdfstring{$N$}{N} limit}
As noted below (\ref{colourexactoneWL}), the expression for the colour-exact Wilson loop obtained from holomorphic linking is not of a closed form, as a 2 Wilson loop correlator appears and the equation cannot be used recursively. This is a general property of the expressions obtained for colour-exact correlators of any number of Wilson loops; an $r$ Wilson loop correlator depends on an $r+1$ Wilson loop correlator. However, such correlators are of a lower MHV degree, so by using the general holomorphic linking expression for a correlator of $r$ Wilson loops, we can generate correlators for any number of Wilson loops (including the simplest case of a single Wilson loop) recursively.

The derivation for the recursion relation for the colour-exact correlator of $r$ Wilson loops,\\
$\langle \mathcal{L}(C_1) \hdots \mathcal{L}(C_r)\rangle$, follows in an analogous way to the single and two Wilson loop calculations in sections \ref{hololinkintro} and \ref{2WLrecursion}. Taking the lines of each Wilson loop to depend on a parameter $t$, the variation of the correlator is given by,
\begin{align}
&\bar{\delta} \langle \mathcal{L}(C_1)\hdots \mathcal{L}(C_r)\rangle = \notag \\
&-\frac{N}{\beta_1 (2 \pi i)^2} \sum_{u=1}^r \sum_{i_u,j_u}  \int  \frac{ds}{s} \wedge\frac{ds'}{s'} \wedge g^* \bar{\delta}^{3|4}(\mathcal{Z},\mathcal{Z'})
\Bigl[ \langle \mathcal{L}(C^\prime_u) \mathcal{L}(C^{\prime \prime}_u) \textstyle\prod_{w\in R\backslash\{ u\}}\displaystyle\mathcal{L}(C_w)\rangle - \frac{\alpha}{N^2} \langle \mathcal{L}(C_1)\hdots \mathcal{L}(C_r)\rangle\Bigr] \notag \\
&-\frac{1}{\beta_1 (2 \pi i)^2 N^2} \sum_{u\neq v}\sum_{i_u,j_v} \int  \frac{ds}{s} \wedge\frac{d\tilde{s}}{\tilde{s}} \wedge g^* \bar{\delta}^{3|4}(\mathcal{Z},\tilde{\mathcal{Z}})\Bigl[ \langle \tr ({\rm Hol}_{\mathcal{Z}}[C_u] {\rm Hol}_{\tilde{\mathcal{Z}}}[C_v])\textstyle\prod_{w\in R \backslash \{u,v\}}\displaystyle \mathcal{L}(C_w)\rangle\notag\\
&\hspace{83mm}- \frac{\alpha}{N} \langle \tr {\rm Hol}_{\mathcal{Z}}[C_u] \tr {\rm Hol}_{\tilde{\mathcal{Z}}}[C_v]\textstyle\prod_{w\in R \backslash \{u,v\}}\displaystyle \mathcal{L}(C_w)\rangle \Bigr] 
\end{align}
where $R=\{1,\hdots,r\}$ are the labels of the $r$ Wilson loops, $i_u$ labels the cusp $i$ of the $u$-th Wilson loop, $i_u\in\{1_u,\hdots,n_u\}$ with $n_u$ the number of cusps on $\mathcal{L}(C_u)$, $\mathcal{Z}\equiv \mathcal{Z}_{X_{i_u,t}}(s), \hspace{1mm}\mathcal{Z^\prime}\equiv \mathcal{Z}_{X_{j_u,t}}(s^\prime)$ and $\mathcal{\tilde{Z}}\equiv \mathcal{Z}_{X_{j_v,t}}(\tilde{s})$. Taking only the line $(\mathcal{Z}_{n_1}\hspace{1mm} \mathcal{Z}_{1_1})$ to be $t$-dependent and applying the same shift as in (\ref{BCFWshift}), we obtain the recursion relation for colour-exact correlators of Wilson loops (which also facilitates the evaluation of a single Wilson loop),
\begin{align}
    &\langle \mathcal{L}(C_1) \hdots \mathcal{L}(C_r)\rangle = \langle \mathcal{L}[1,\ldots,n_1-1] \textstyle\prod_{w\in R\backslash \{1\} }\displaystyle \mathcal{L}(C_w)\rangle) \label{bcfwcolourexact}\\
    & +\frac{N}{2C_F} \sum_{j_1=3}^{n_1-2} [n_1-1,n_1,1_1,j_1-1,j_1] \notag\\
    &\hspace{10mm}\times \Bigl(\langle \mathcal{L}[1_1,\ldots,j_1-1,I_{j_1}]\mathcal{L}[I_{j_1},j_1,\ldots,\hat{n}_{1,j_1}] \textstyle\prod_{w\in R\backslash \{1\}}\displaystyle \mathcal{L}(C_w)\rangle-\frac{\alpha}{N^2}\langle \mathcal{L}(C_1) \hdots \mathcal{L}(C_r)\rangle \Bigr) \notag \\
    &  +\frac{N}{2C_F}\frac{1}{N^2} \sum_{v \in R\backslash\{1\}}\sum_{j_v=1_v}^{n_v} [n_1-1,n_1,1_1,j_v-1,j_v]\notag\\
    &\hspace{10mm}\times\Bigl( \langle \mathcal{L}[1_1,\ldots,\hat{n}_{1,j_v},I_{j_v},j_v,\ldots,j_v-1,I_{j_v}] \textstyle \prod_{w \in R\backslash \{1,v\}} \displaystyle \mathcal{L}(C_w)\rangle- \alpha \langle \mathcal{L}[1_1,\ldots,\hat{n}_{1,j_v}]  \textstyle\prod_{w\in R\backslash \{1\}}\displaystyle \mathcal{L}(C_w)
    \rangle\Bigr) \notag
\end{align}
where
\begin{align}
    &\hat{Z}_{n_1,j_v}\equiv (n_1-1 \hspace{1mm} n_1) \cap (1_1\hspace{1mm} j_v-1 \hspace{1mm} j_v) = Z_{n_1-1}\langle n_1 \hspace{1mm} 1_1\hspace{1mm} j_v-1 \hspace{1mm} j_v\rangle- Z_{n_1}\langle n_1-1 \hspace{1mm} 1_1 \hspace{1mm} j_v -1 \hspace{1mm} j_v\rangle \notag\\
    &Z_{I_{j_v}}\equiv (j_v-1 \hspace{1mm} j_v) \cap (n_1-1 \hspace{1mm} n_1 \hspace{1mm} 1_1) = Z_{j_v-1}\langle j_v \hspace{1mm} n_1-1 \hspace{1mm} n_1 \hspace{1mm} 1_1\rangle- Z_{j_v}\langle j_v-1 \hspace{1mm} n_1-1 \hspace{1mm} n_1 \hspace{1mm} 1_1\rangle 
\end{align}
The $r=1$ case is given explicitly in (\ref{colourexactoneWL0}). 

\subsection{Example: N\texorpdfstring{\(^2\)}{2}MHV hexagon}

As an instructive example of how this recursion works, we can compute the colour-exact tree-level contribution to the hexagon $\langle \mathcal{L}[1,\hdots,6]\rangle$ at N\(^2\)MHV for U$(N)$ from the recursion relation. Using (\ref{bcfwcolourexact}), we have
\begin{align}
    \langle \mathcal{L}[1,\hdots,6]\rangle^{(0,2)} = & \langle \mathcal{L}[1,\hdots,5]\rangle^{(0,2)}\notag\\
    &+\frac{N}{2C_F}( [5,6,1,2,3]\langle \mathcal{L}[1,2,I_3]\mathcal{L}[I_3,3,4,5,\hat{6}_3]\rangle^{(0,1)}\notag\\
    &\hspace{16.5mm}+ [5,6,1,3,4]\langle \mathcal{L}[1,2,3,I_4]\mathcal{L}[I_4,4,5,\hat{6}_4]\rangle^{(0,1)} )
    \label{firstlevel}
\end{align}
where $\langle \mathcal{L}(C_1) \hdots \mathcal{L}(C_r)\rangle^{(l,k)}$ refers to the $O(g^{2l})$ N$^k$MHV correlator, $I_j=(j-1 \hspace{1mm} j)\cap (1 \hspace{1mm} 5 \hspace{1mm} 6)$ and $\hat{6}_j=(5 \hspace{1mm} 6)\cap (1 \hspace{1mm} j-1 \hspace{1mm} j)$. $\langle \mathcal{L}[1,\hdots,5]\rangle^{(0,2)}=0$ and using (\ref{bcfwcolourexact}) again for the first correlator, we get
\begin{align}
    \langle \mathcal{L}[1,2,I_3]\mathcal{L}[I_3,3,4,5,\hat{6}_3]\rangle^{(0,1)}& =\langle \mathcal{L}[1,2]\mathcal{L}[I_3,3,4,5,\hat{6}_3]\rangle^{(0,1)}+\frac{1}{2N C_F}\sum_{j\in \{4,5,\hat{6}_3\}}[2,I_3,1,j-1,j]\notag\\
    & =\langle\mathcal{L}[I_3,3,4,5,\hat{6}_3]\rangle^{(0,1)}+\frac{1}{2N C_F}\sum_{j\in \{5,\hat{6}_3\}}[2,I_3,1,j-1,j]\notag\\
    & =\frac{N}{2C_F}[5,\hat{6}_3,I_3,3,4]+\frac{1}{2N C_F}\sum_{j\in \{5,\hat{6}_3\}}[2,I_3,1,j-1,j]
\end{align}
where (\ref{bcfwcolourexact}) was used again in the last line. We use recursion to compute the second correlator in (\ref{firstlevel}),
\begin{align}
    \langle \mathcal{L}[1,2,3,4]\mathcal{L}[I_4,4,5,\hat{6}_4]\rangle^{(0,1)}& =\langle \mathcal{L}[1,2,3]\mathcal{L}[I_4,4,5,\hat{6}_4]\rangle^{(0,1)}+\frac{1}{2N C_F}[3,I_4,1,5,\hat{6}_4]\notag\\
    &=\frac{1}{2N C_F }\sum_{j\in \{I_4,4,5,\hat{6}_4\}}[2,3,1,j-1,j]+\frac{1}{2N C_F}[3,I_4,1,5,\hat{6}_4]\notag\\
\end{align}
Plugging these two expressions into (\ref{firstlevel}), we have
\begin{align}
    \langle \mathcal{L}[1,\hdots,6]\rangle^{(0,2)}= & \frac{N^2}{4C_F^2}\Bigl([5,6,1,2,3]([5,\hat{6}_3,I_3,3,4]+\frac{1}{N^2}\sum_{j\in \{5,\hat{6}_3\}}[2,I_3,1,j-1,j])\notag\\
    &\hspace{6mm}+\frac{1}{N^2}[5,6,1,3,4]([3,I_4,1,5,\hat{6}_4]+\sum_{j\in\{I_4,4,5,\hat{6}_4\}}[2,3,1,j-1,j])\Bigr)
\end{align}
Checking this expression on different chi monomials on random numerics, we find a match with the diagrammatic calculation. By making use of (\ref{bcfwcolourexact}) for $r=1$ and $r=2$, we were able to find the colour exact expression for N$^2$MHV hexagon at tree-level for U$(N)$. As we are at maximal MHV degree, the colour exact hexagon can only differ from the planar expression by a bosonic factor, which depends in a simple way on conformally invariant cross ratios. By probing the expression on different numerics, we find that they are related by,
\begin{equation}
    \langle \mathcal{L}[1,\hdots,6] \rangle^{(0,2)} = \left(1+\frac{1}{N^2}(u_1+u_2+u_3-1)\right)\langle \mathcal{L}[1,\hdots,6]\rangle^{(0,2),p},
\end{equation}
where
\begin{equation}
    u_1=\frac{x_{13}^2x_{46}^2}{x_{36}^2x_{41}^2}\hspace{5mm}u_2=\frac{x_{15}^2x_{24}^2}{x_{14}^2x_{25}^2}\hspace{5mm} u_3=\frac{x_{26}^2 x_{35}^2}{x_{25}^2x_{36}^2}.
\end{equation}
and $p$ means the planar part.

We can compute general colour-exact Wilson loop correlators in a completely analogous way by implementing the recursion relation on a computer. Note that since every correlator which features in the BCFW recursion relation is either of one point lower (the first, boundary term) or enters at a lower MHV degree due to the R-invariant pre-multiplying the other terms, the recursion will always terminate. Note also that the same colour-exact recursion can be performed for loop integrands. Although the `forward limit' term in this case will feature a Wilson loop correlator at MHV degree one \emph{higher} than the target expression (due to the fermionic integration to be performed on this term), the loop level correlator which enters is at one loop order lower and so we can subsequently construct that object recursively until the procedure terminates in the same way. 

\section{Conclusions}
We have demonstrated that holomorphic linking gives rise to a BCFW-like recursion relation for the correlators of multiple light-like Wilson loop operators, which generalises the BCFW relation familiar from the study of a single Wilson loop's expectation value (or, equivalently, the dual scattering amplitude). We have verified these recursion relations in a number of cases, both analytically and numerically, and have demonstrated that they allow colour-exact Wilson loop correlators to be computed recursively. 

It would be interesting to further study the practical usefulness of these recursion relations for computation. We already made use of them in \cite{Drummond:2026lvq} to simplify the verification of the \(\bar{Q}\)-equation and in particular to simplify the action of the `collinear integral' on tree-level Wilson loop correlators. Although the MHV/twistor Wilson loop diagram rules are in practice very simple to implement, at high multiplicity and/or MHV degree the number of diagrams grows very rapidly and the BCFW recursions which we have presented here should allow for the generation of more compact expressions, albeit featuring e.g. more complicated dependence on the kinematics per term. It would be interesting to study the precise growth rate in the number of terms in either representation, and to experiment with e.g. applying the recursion all the way down versus, say, stopping at the simplified cases which factorise (which are N\(^2\)MHV tree, NMHV one-loop, and MHV two-loops, in the SU($N$) theory). Relatedly, it would be instructive and useful to automate these recursion relations in a computer algebra system. 

One pressing line of enquiry is whether the recursion relations which we have presented here hint at a description of Wilson loop correlators in terms of positive geometries (or some generalisation thereof). Since the ordinary BCFW recursion relations provide a triangulation of the amplituhedron, a natural question is whether these BCFW-like relations can play a similar role for a generalisation of the amplituhedron to the case of multiple loop operators. It would also be interesting to investigate whether the BCFW representation for Wilson loop correlators admit any notion of cluster compatibility for the poles common to each term, as is the case for a single Wilson loop \cite{Drummond:2018dfd}. We defer exploration of these points to further work. 

We also find it interesting to note that the recursion relations we present here make very explicit the connection between a colour-exact correlator of \(r\)-Wilson loops, and correlators of \(r+1\)-Wilson loops (at lower MHV degree). For instance, a colour-exact \emph{single} Wilson loop's expectation value is related to the correlator of two Wilson loops at MHV degree one lower, which we can further decompose into higher but simpler correlators using the recursion relations presented here. This suggests a deep connection between the non-planar corrections to a single Wilson loop's expectation value, and the correlation functions of multiple Wilson loops. It would be interesting to explore whether, as a consequence, the familiar formulae for the planar one-loop leading singularities of a single Wilson loop's expectation value admit non-planar generalisations, for instance in terms of correlators, rather than products, of loop operators at tree level. It may be possible to address this question by adapting the arguments presented in \cite{Drummond:2026gpt}, and we hope to return to this point in future work.

\section*{Acknowledgements}

All authors are supported by the STFC consolidated grant ST/X000583/1.

\appendix

\end{document}